\documentclass{lmcs}
\pdfoutput=1

\usepackage{silence}
\usepackage{lastpage}
\lmcsdoi{16}{1}{30}
\lmcsheading{}{\pageref{LastPage}}{}{}%
{Nov.~15,~2017}{Mar.~10,~2020}{}

\author[M.~Rennela]{Mathys Rennela\rsuper{a}}
\address{\lsuper{a}Radboud University\\ Nijmegen, The Netherlands}
\email{mathys.rennela@gmail.com}

\author[S.~Staton]{Sam Staton\rsuper{b}}
\address{\lsuper{b}Oxford University\\ Oxford, The United Kingdom}
\email{sam.staton@cs.ox.ac.uk}


\title[Classical control and quantum circuits in enriched category theory]{Classical control, quantum circuits and linear logic \texorpdfstring{\\}{} in enriched category theory}

\RequirePackage[l2tabu, orthodox]{nag}

\usepackage{amsthm,amsmath,amssymb}
\usepackage{mathtools}
\usepackage{bm}
\usepackage{cite}
\usepackage[all,cmtip]{xy}
\usepackage{appendix}

\usepackage{stmaryrd}
\usepackage{comment}
\usepackage{xspace}
\usepackage{subfig}
\usepackage{yfonts}
\usepackage{microtype} 
\usepackage{tikz}						
\usepackage{wrapfig}
\usetikzlibrary{arrows,shapes,automata,backgrounds,positioning,petri,topaths}

\usepackage{float}
\floatstyle{boxed}
\restylefloat{figure}

\newcommand{\defeq}{\stackrel{\text{def}}=}
\newcommand{\abs}[1]{\left| #1 \right|}

\newcommand{\C}{\mathbb{C}}

\newcommand{\N}{\mathbb{N}}

\newcommand{\unit}{[0,1]}

\newcommand{\cat}[1]{\mathbf{#1}}
\newcommand{\Conv}{\mathbf{Conv}}

\newcommand{\Dcpo}{\mathbf{Dcpo}}
\newcommand{\DcpoS}{\ensuremath{\Dcpo_{\perp !}}}

\newcommand{\DM}{\mathcal{D}_{= 1}}

\newcommand{\EMod}{\mathbf{EMod}}

\newcommand{\dEMod}{\mathbf{dEMod}}

\newcommand{\hsEMod}{\hyperstonean\EMod}

\newcommand{\Fd}{\mathbf{Fd}}

\newcommand{\WStar}{\mathbf{W^*\text{-}Alg}}
\newcommand{\CStar}{\mathbf{C^*\text{-}Alg}}

\newcommand{\CPU}{\CStar_\mathrm{CPU}}
\newcommand{\CPSU}{\CStar_\mathrm{CPSU}}

\newcommand{\wCPSU}{\WStar_\mathrm{CPSU}}

\newcommand{\opp}[1]{#1^\mathbf{op}}

\newcommand{\tensor}{\overline{\otimes}}

\newcommand{\new}{\operatorname{new}}
\newcommand{\meas}{\operatorname{meas}}

\newcommand{\mat}[1]{M_{#1}}

\newcommand{\CC}{\mathbb{C}}
\newcommand{\myparagraph}[1]{\noindent\textit{#1.}}
\newcommand{\FdCPU}{\Fd\CPU}

\newcommand{\hide}[1]{}

\newcommand{\denot}[1]{[\![#1]\!]}

\newdimen\proofrulebreadth \proofrulebreadth=.05em
\newdimen\proofdotseparation \proofdotseparation=1.25ex
\newdimen\proofrulebaseline \proofrulebaseline=2ex
\newcount\proofdotnumber \proofdotnumber=3
\let\then\relax
\def\hfi{\hskip0pt plus.0001fil}
\mathchardef\squigto="3A3B
\newif\ifinsideprooftree\insideprooftreefalse
\newif\ifonleftofproofrule\onleftofproofrulefalse
\newif\ifproofdots\proofdotsfalse
\newif\ifdoubleproof\doubleprooffalse
\let\wereinproofbit\relax
\newdimen\shortenproofleft
\newdimen\shortenproofright
\newdimen\proofbelowshift
\newbox\proofabove
\newbox\proofbelow
\newbox\proofrulename
\def\shiftproofbelow{\let\next\relax\afterassignment\setshiftproofbelow\dimen0 }
\def\shiftproofbelowneg{\def\next{\multiply\dimen0 by-1 }%
\afterassignment\setshiftproofbelow\dimen0 }
\def\setshiftproofbelow{\next\proofbelowshift=\dimen0 }
\def\setproofrulebreadth{\proofrulebreadth}

\def\prooftree{
\ifnum  \lastpenalty=1
\then   \unpenalty
\else   \onleftofproofrulefalse
\fi
\ifonleftofproofrule
\else   \ifinsideprooftree
        \then   \hskip.5em plus1fil
        \fi
\fi
\bgroup
\setbox\proofbelow=\hbox{}\setbox\proofrulename=\hbox{}%
\let\justifies\proofover\let\leadsto\proofoverdots\let\Justifies\proofoverdbl
\let\using\proofusing\let\[\prooftree
\ifinsideprooftree\let\]\endprooftree\fi
\proofdotsfalse\doubleprooffalse
\let\thickness\setproofrulebreadth
\let\shiftright\shiftproofbelow \let\shift\shiftproofbelow
\let\shiftleft\shiftproofbelowneg
\let\ifwasinsideprooftree\ifinsideprooftree
\insideprooftreetrue
\setbox\proofabove=\hbox\bgroup$\displaystyle 
\let\wereinproofbit\prooftree
\shortenproofleft=0pt \shortenproofright=0pt \proofbelowshift=0pt
\onleftofproofruletrue\penalty1
}

\def\eproofbit{
\ifx    \wereinproofbit\prooftree
\then   \ifcase \lastpenalty
        \then   \shortenproofright=0pt  
        \or     \unpenalty\hfil         
        \or     \unpenalty\unskip       
        \else   \shortenproofright=0pt  
        \fi
\fi
\global\dimen0=\shortenproofleft
\global\dimen1=\shortenproofright
\global\dimen2=\proofrulebreadth
\global\dimen3=\proofbelowshift
\global\dimen4=\proofdotseparation
\global\count255=\proofdotnumber
$\egroup  
\shortenproofleft=\dimen0
\shortenproofright=\dimen1
\proofrulebreadth=\dimen2
\proofbelowshift=\dimen3
\proofdotseparation=\dimen4
\proofdotnumber=\count255
}

\def\proofover{
\eproofbit 
\setbox\proofbelow=\hbox\bgroup 
\let\wereinproofbit\proofover
$\displaystyle
}%
\def\proofoverdbl{
\eproofbit 
\doubleprooftrue
\setbox\proofbelow=\hbox\bgroup 
\let\wereinproofbit\proofoverdbl
$\displaystyle
}%
\def\proofoverdots{
\eproofbit 
\proofdotstrue
\setbox\proofbelow=\hbox\bgroup 
\let\wereinproofbit\proofoverdots
$\displaystyle
}%
\def\proofusing{
\eproofbit 
\setbox\proofrulename=\hbox\bgroup 
\let\wereinproofbit\proofusing
\kern0.3em$
}

\def\endprooftree{
\eproofbit 
  \dimen5 =0pt
\dimen0=\wd\proofabove \advance\dimen0-\shortenproofleft
\advance\dimen0-\shortenproofright
\dimen1=.5\dimen0 \advance\dimen1-.5\wd\proofbelow
\dimen4=\dimen1
\advance\dimen1\proofbelowshift \advance\dimen4-\proofbelowshift
\ifdim  \dimen1<0pt
\then   \advance\shortenproofleft\dimen1
        \advance\dimen0-\dimen1
        \dimen1=0pt
        \ifdim  \shortenproofleft<0pt
        \then   \setbox\proofabove=\hbox{%
                        \kern-\shortenproofleft\unhbox\proofabove}%
                \shortenproofleft=0pt
        \fi
\fi
\ifdim  \dimen4<0pt
\then   \advance\shortenproofright\dimen4
        \advance\dimen0-\dimen4
        \dimen4=0pt
\fi
\ifdim  \shortenproofright<\wd\proofrulename
\then   \shortenproofright=\wd\proofrulename
\fi
\dimen2=\shortenproofleft \advance\dimen2 by\dimen1
\dimen3=\shortenproofright\advance\dimen3 by\dimen4
\ifproofdots
\then
        \dimen6=\shortenproofleft \advance\dimen6 .5\dimen0
        \setbox1=\vbox to\proofdotseparation{\vss\hbox{$\cdot$}\vss}%
        \setbox0=\hbox{%
                \advance\dimen6-.5\wd1
                \kern\dimen6
                $\vcenter to\proofdotnumber\proofdotseparation
                        {\leaders\box1\vfill}$%
                \unhbox\proofrulename}%
\else   \dimen6=\fontdimen22\the\textfont2 
        \dimen7=\dimen6
        \advance\dimen6by.5\proofrulebreadth
        \advance\dimen7by-.5\proofrulebreadth
        \setbox0=\hbox{%
                \kern\shortenproofleft
                \ifdoubleproof
                \then   \hbox to\dimen0{%
                        $\mathsurround0pt\mathord=\mkern-6mu%
                        \cleaders\hbox{$\mkern-2mu=\mkern-2mu$}\hfill
                        \mkern-6mu\mathord=$}%
                \else   \vrule height\dimen6 depth-\dimen7 width\dimen0
                \fi
                \unhbox\proofrulename}%
        \ht0=\dimen6 \dp0=-\dimen7
\fi
\let\doll\relax
\ifwasinsideprooftree
\then   \let\VBOX\vbox
\else   \ifmmode\else$\let\doll=$\fi
        \let\VBOX\vcenter
\fi
\VBOX   {\baselineskip\proofrulebaseline \lineskip.2ex
        \expandafter\lineskiplimit\ifproofdots0ex\else-0.6ex\fi
        \hbox   spread\dimen5   {\hfi\unhbox\proofabove\hfi}%
        \hbox{\box0}%
        \hbox   {\kern\dimen2 \box\proofbelow}}\doll%
\global\dimen2=\dimen2
\global\dimen3=\dimen3
\egroup 
\ifonleftofproofrule
\then   \shortenproofleft=\dimen2
\fi
\shortenproofright=\dimen3
\onleftofproofrulefalse
\ifinsideprooftree
\then   \hskip.5em plus 1fil \penalty2
\fi
}

Date: Tue, 19 May 1998 16:45:32 +0100
From: Simon Gay <simon@dcs.rhbnc.ac.uk>

I've got another problem when combining
your packages with elsart.cls. The code

\documentclass{elsart}
\input{prooftree}
\input{diagrams}
\begin{document}

\[
\begin{prooftree}
A \rTo^{f} B
\justifies
p \rTo^{(a,c)} q
\end{prooftree}
\]

\end{document}

doesn't leave enough space below the line in the proof tree, so that
the (a,c) label on the lower arrow runs into the line. It's fine with
article.cls. 

\input{Qcircuit}

\newcommand{\kindoftheorem}{Theorem}
\newtheorem*{theoremcustom}{\kindoftheorem}

\newcommand{\Ddirinf}{\mathcal{D}_\infty^\uparrow}
\newcommand{\Ddir}{\mathcal{D}^\uparrow}
\renewcommand{\dEMod}{\mathbf{dc}\EMod}
\renewcommand{\hsEMod}{\mathbf{hs}\EMod}
\newcommand{\hsConv}{\mathbf{hs}\Conv}
\renewcommand{\DM}{\mathcal{D}}

\newcommand{\newfun}{\textsc{new}}
\newcommand{\applyfun}{\textsc{apply}}
\newcommand{\measfun}{\textsc{meas}}
\newcommand{\qlist}{\op{qlist}}
\newcommand{\cointoss}{\op{cointoss}}
\newcommand{\entangle}{\op{entangle}}
\newcommand{\ketbf}[1]{\bm{\ket{#1}}}
\newcommand{\succfun}{\textsc{succ}}
\newcommand{\predfun}{\textsc{pred}}
\newcommand{\splitop}{\op{split}}
\newcommand{\inop}{\op{in}}
\newcommand{\asop}{\op{as}}
\newcommand{\letrecop}{\op{letrec}}
\newcommand{\introop}{\op{intro}}
\newcommand{\elimop}{\op{elim}}

\newcommand{\yoneda}{\mathbf{y}}

\newcommand{\Set}{\cat{Set}}
\newcommand{\Lawv}[1]{\mathbb{L}_{#1}}
\newcommand{\LawvConv}{\Lawv{}}
\newcommand{\LawvSConv}{\Lawv{\leq 1}} 

\newcommand{\cwA}{\mathbf{a}}
\newcommand{\cwB}{\mathbf{b}}
\newcommand{\qwA}{\alpha}
\newcommand{\qwB}{\beta}
\newcommand{\Circ}{\mathrm{Circ}}
\newcommand{\tbool}{\mathrm{bit}}
\newcommand{\tqubit}{\mathrm{qubit}}
\newcommand{\classicalize}[1]{\overline{#1}}
\newcommand{\monad} T
\newcommand{\ireturn}[1]{\ensuremath{\mathbf{return}(#1)}}
\newcommand{\iletin}[3]{\ensuremath{\mathbf{let}\,#2=#1\,\mathbf{in}\,#3}}
\newcommand{\newq}{\mathrm{new}}
\newcommand{\tid}{\text{id}}
\newcommand{\tunbox}{\textbf{unbox }}
\newcommand{\tbox}{\textbf{box }}
\newcommand{\tgate}{\textbf{gate }}
\newcommand{\toutput}{\textbf{output }}
\newcommand{\tin}{\text{in}}
\newcommand{\tqlist}{\text{QList}}
\newcommand{\tqnat}{\text{QNat}}
\newcommand{\tnat}{\text{Nat}}

\renewcommand{\tensor}{\otimes}
\renewcommand{\opp}[1]{#1^{\mathrm{op}}}
\newcommand{\homDay}[2]{[#1,#2]_\text{Day}}
\newcommand{\Day}{\otimes_\text{Day}}

\newcommand{\FDCP}{\Fd\CStar_{\mathrm{CP}}}
\newcommand{\CFDCP}{\Fd\mathbf{C}\CStar_{\mathrm{CP}}}
\newcommand{\CFDCPU}{\Fd\mathbf{C}\CStar_{\mathrm{CPU}}}
\newcommand{\CFDCPSU}{\Fd\mathbf{C}\CStar_{\mathrm{CPSU}}}

\newcommand{\mmm}[1]{\textcolor{blue}{\textbf{Mathys:~#1}}}
\newcommand{\sss}[1]{\textcolor{red}{\textbf{Sam:~#1}}}

\newcommand{\colim}{\mathrm{colim}}
\newcommand{\op}{^{\mathrm{op}}}

\usepackage{listings}
\lstdefinestyle{customc}{
  belowcaptionskip=1\baselineskip,
  breaklines=true,
  frame=L,
  xleftmargin=\parindent,
  language=ML,
  showstringspaces=false,
  basicstyle=\small\ttfamily,
  tabsize=2,
  keywordstyle=\bfseries\color{black},
  commentstyle=\itshape\color{purple!40!black},
  identifierstyle=\color{darkgray!99!black},
  breakatwhitespace=false,
  breaklines=true,
  captionpos=b,
  frame=none,
  keepspaces=true,
  escapeinside={(*}{*)}, 
  morekeywords={box,case,of,in1,in2,then,init,output,gate,unbox,run,lift},           
}
\lstset{escapechar=@,style=customc}

\begin{document}
\maketitle

\begin{abstract}
We describe categorical models of a circuit-based (quantum) functional programming language. We show that enriched categories play a crucial role. Following earlier work on QWire by Paykin et al., we consider both a simple first-order linear language for circuits, and a more powerful host language, such that the circuit language is embedded inside the host language. Our categorical semantics for the host language is standard, and involves cartesian closed categories and monads. We interpret the circuit language not in an ordinary category, but in a category that is enriched in the host category. We show that this
structure is also related to linear/non-linear models. As an extended example, we recall an earlier result that the category of W*-algebras is dcpo-enriched, and we use this model to extend the circuit language with some recursive types.
\end{abstract}

\section*{Introduction}

\begin{wrapfigure}{r}{0.3\textwidth}
\centering
\begin{tikzpicture}
\node (classical) {\textbf{Classical}};
\node (quantum) [right=1mm of classical] {\textbf{Quantum}};
\draw[->] (classical) edge [bend left] node [left,above] {control} (quantum);
\draw[->] (quantum) edge [bend left] node [right, below] {measurements} (classical);
\end{tikzpicture}
\end{wrapfigure}

One of the subtle points about quantum computation is the interaction between
classical control flow and quantum operations.
One can measure a qubit, destroying the qubit but producing a classical bit;
this classical bit can then be used to decide whether to apply quantum rotations to other qubits.
This kind of classical control can be neatly described in quantum circuits, for example when one uses the measurement outcome of a qubit $a$ to conditionally perform a gate $X$ on a qubit $b$:
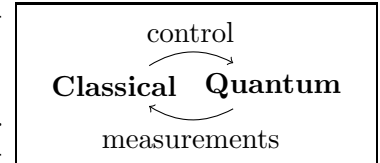
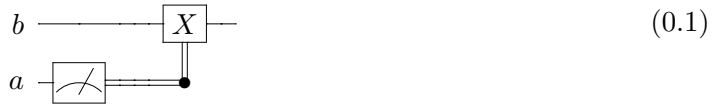
\begin{equation}\label{qcircuit-example}
\Qcircuit @C=.5em @R=.7em {
\lstick b& \qw & \qw&\qw&\qw& \gate{X}& \qw & \qw
\\
\lstick a & \meter & \cw & \cw & \cw & \control{-1}\cwx[-1]\cw
}\end{equation}

This can be understood
semantically in terms of mixed states, density matrices, and completely positive maps. However, high level languages have more elaborate data structures than bits: they have higher order functions and mixed variance recursive types, and associated with these are elaborate control structures such as higher order recursive functions. These are important, paradigmatic ways of structuring programs.

How should these high level features be integrated with quantum computation?
One option is to build a semantic domain that accommodates both quantum computation and higher order features. This is an aim of some categorical semantics of the quantum lambda calculus~\cite{pagani-selinger-valiron-popl14,malherbe-scott-selinger} and of prior work of the authors~\cite{rennela-mfps30,rennela-staton-mfps31}. This is a fascinating direction, and sheds light, for example, on the structure of the quantum teleportation algorithm (e.g.~\cite[Example~6]{pagani-selinger-valiron-popl14}). However, the general connection between physics and higher-order quantum functions is yet unclear. Although some recent progress has been made~\cite{kissinger-uijlen-lics17}, it is still unclear whether higher-order quantum functions of this kind are useful for quantum algorithms.

Another approach is to understand a high level quantum programming language as an ordinary higher-order functional language with extra features for building and running quantum circuits. In this setting, quantum circuits form a first-order embedded domain specific language within a conventional higher order language. This fits the current state-of-the-art in interfaces to quantum hardware, and is the basis of the languages Quipper~\cite{quipper} and LiQUi$|\rangle$~\cite{liquid}.
This is the approach that we study in this paper.

\subsection*{Embedded languages and enriched categories}
Our work revolves around a new calculus that we call `EWire' (\S\ref{sec:qwire}). It is a minor generalization of the QWire language~\cite{qwire,randqwire}. QWire idealizes some aspects of the architecture of Quipper and LiQUi$|\rangle$.
The idea is that we deal with a host language separated from an embedded circuit language.
\begin{itemize}
\item The circuit language is a first order typed language.
The types, called `wire types', include a type for qubits. The wire type system is \textit{linear} to
accommodate the fact that qubits cannot be duplicated.
\item The host language is a higher order language. The types of the host language do not
include the wire types, there is not a type of qubits, and it is not a linear type system.
However, there is a special host type $\mathrm{Circ}(W_1,W_2)$
associated to any pair of wire types $W_1$ and $W_2$, whose inhabitants are the circuits with inputs of type $W_1$ and
outputs of type $W_2$.
\end{itemize}

\noindent
Let us describe the circuit language in a nutshell: the very simple circuit~\eqref{qcircuit-example} corresponds to the instruction below~\eqref{eqn:example} in the circuit language. Given two qubits $a$ and $b$, it measures the qubit $a$, stores the result in a bit $x$ which is later used in the application of the classical-controlled-$X$ gate and discards the bit $x$, then outputs the resulting qubit $y$.
\begin{align}
\notag -;a,b:\mathrm{qubit}\ \vdash\  C\ \  \defeq \ \
 &x \leftarrow \textbf{gate } \textbf{meas } a;
 (x,y) \leftarrow \textbf{gate } (\textbf{bit-control } X)\, (x,b);\\
 &() \leftarrow \textbf{gate } \textbf{discard } x;
 \textbf{output } y
 \ \qquad :\ \mathrm{qubit}
\label{eqn:example}
\end{align}
The interface between the host language and the circuit language is set up in terms of boxing and unboxing. For example, the instruction
\begin{equation}\label{eqn:boxedexample}
t\ \  \defeq \ \  \textbf{box } (a,b) \Rightarrow C (a,b)\qquad \text{(where $C$ is as in~\eqref{eqn:example})}
\end{equation}
creates a closed term of type $\text{Circ(qubit$\otimes$qubit,qubit)}$ in the host language. We recover the instruction $C$ in the circuit language~\eqref{eqn:example} from the boxed expression $t$ in the host language~\eqref{eqn:boxedexample} by using the instruction $\textbf{unbox } t\, w$ for some fresh wire $w$ of type qubit. 

Also, it is possible to write a program that
composes two circuits $C_1$ and $C_2$ with the right input/output types, for example:
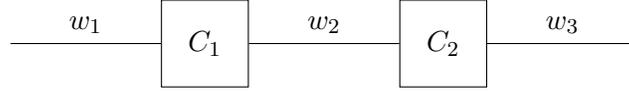
\tikzstyle{block} = [draw, rectangle, minimum height=3em, minimum width=3em]
\tikzstyle{virtual} = [coordinate]
\begin{center}
\begin{tikzpicture}[>=stealth,auto, node distance=2cm]
    \node [block] (C1) {$C_1$};
    \node [block, right=of C1] (C2) {$C_2$};

    \node [virtual, left=of C1] (w1)     {};
    \node [virtual, right=of C1] (w2)    {};
    \node [virtual, right=of C2] (w3)    {};

    \draw [-] (w1) -- node {$w_1$} (C1);
    \draw [-] (C1) -- node {$w_2$} (w2);
    \draw [-] (C2) -- node {$w_3$}(w3);
\end{tikzpicture}
\end{center}
This is a program
\[
\text{comp }\ \defeq\
\lambda (C_1,C_2).\
\textbf{box } w_1 \Rightarrow
\big(w_2 \leftarrow \textbf{unbox } C_1 w_1;
w_3\leftarrow \textbf{unbox } C_2 w_2 ; \textbf{output }w_3\big)
\]
in the host language, associated with the type
\begin{equation}\label{eqn:composition}
\text{comp}\ : \ \mathrm{Circ}(W_1,W_2)\times \mathrm{Circ}(W_2,W_3)\to \mathrm{Circ}(W_1,W_3)
\end{equation}
where $W_i$ is the type of the wire $w_i$ for $i \in \{1,2,3\}$.

Now, recall the idea of an enriched category, which is informally a
category such that the morphisms from $A$ to $B$ form an object of another category.
In Section~\ref{sec:cat-qwire}, once we conceptualize types as objects and terms as morphisms,
we show that \textbf{\emph{the embedding of the circuit language in the host language is an
instance of enriched category theory}}:
the circuits (morphisms) between wire types (objects) form a type (object) of the host language.
The host composition term in~\eqref{eqn:composition} is precisely composition in the sense of
enriched categories.

This enriched categorical framework is then incorporated in the definition of categorical models of EWire (Definition~\ref{def:model-QWire}), which is associated to the following proposition, building on a partial normalization procedure proposed in~\cite{qwire}
for QWire expressions, which is shown to be sound with respect to a semantics in completely positive maps~\cite[App.~A--B]{qwire}.

\renewcommand{\kindoftheorem}{Proposition~\ref{prop:soundness}}
\begin{theoremcustom}[paraphrased]
The program equations that are used to justify the steps in the normalization procedure in~\cite{qwire} are sound in any EWire model.
\end{theoremcustom}

The idea of denoting a language for circuits within a symmetric monoidal category is an established idea (e.g.~\cite{ghica-circuits}). The novel point here is enrichment.

For a simple version of the model, wire types are understood as finite-dimensional C*-algebras,
and circuits are completely positive unital maps --- the accepted model of quantum computation.
Host types are interpreted as sets, and the type of all circuits is interpreted simply as the set of all circuits.
The category of sets supports higher order functions, which shows that it is consistent for the host language to have higher order functions.

\renewcommand{\kindoftheorem}{Proposition~\ref{prop:B2:fdcpu-model}}
\begin{theoremcustom}[paraphrased]
Taking finite-dimensional C*-algebras and completely positive unital maps as a categorical model of the circuit language yields a categorical model of QWire in which the types qubit and bit are respectively interpreted by the C*-algebras $M_2$ and $\CC \oplus \CC$.
\end{theoremcustom}

As with any higher order language, the set-theoretic model is not sufficient to fully understand the nature of higher order functions.
We envisage that other semantic models (e.g.~based on game semantics) will also fit the same framework of enriched categories, so that our categorical framework provides
a sound description of the basic program equivalences that should hold in all models.
These equivalences play the same role that the $\beta$ and $\eta$ equivalences play in the pure lambda calculus.
In other words, we are developing a basic type theory for quantum computation.

\subsection*{Linear/non-linear models}
Quantum resources cannot be duplicated or discarded, where\-as classical resources can.
This suggests a connection to linear logic~\cite{girard-LL}, as many others have observed~\cite{hasuo-hoshino,rios-selinger-qpl17,ross-phd-thesis,malherbe-scott-selinger,lindenhovius-mislove-zamdzhiev,pagani-selinger-valiron-popl14,geometry-parallelism}. In fact, enriched category theory is closely related to linear logic through the
following observation  (see also~\cite{enriched-effect-calculus,mellies-enriched-adjunctions,ms-state}).
\renewcommand{\kindoftheorem}{Lemma~\ref{lemma:enriched-lnl}}
\begin{theoremcustom}[paraphrased]
The following data are equivalent:
\begin{itemize}
\item A symmetric monoidal closed category enriched in a cartesian closed category, with copowers.
\item A`linear/non-linear' (LNL) model in the sense of Benton~\cite{benton}: a symmetric monoidal adjunction between a cartesian closed category and a symmetric monoidal closed category.
\end{itemize}
\end{theoremcustom}

\noindent
Thus, from the semantic perspective, enriched category theory is a way to study fragments and variations of linear logic.
We show that every LNL model, with some minor extra structure, induces an EWire model (see Proposition~\ref{prop:Lwire-to-Ewire}). We call these structures LNL-EWire models.

So LNL-EWire extends EWire with some additional connectives. However, although these additional connectives have an established type theoretic syntax, their
meaning from the perspective of quantum physics is less clear at present.
For example, LNL assumes a first class linear function space, but categories of completely positive unital maps (or CPTP maps) are
typically not monoidal closed.
Nonetheless, there is a semantic way to understand their relevance, through the following theorem, which
is proved using a variation on Day's construction~\cite{day-convolution}, following other recent work~\cite{rios-selinger-qpl17,malherbe-scott-selinger,hearn-bunched-typing}.
\renewcommand{\kindoftheorem}{Theorem~\ref{th:Ewire-to-Lwire}}
\begin{theoremcustom}[paraphrased]Every EWire model (with a locally presentable base) embeds in an LNL-EWire model. \end{theoremcustom}
\subsection*{Recursive types and recursive terms}
Within our semantic model based on enriched categories, we can freely accommodate various additional features in the host language, while keeping the
circuit language the same.
For example, we could add recursion to the host language, to model the idea of repeatedly trying quantum experiments, or
recursive types, to model arbitrary data types.
This can be shown to be consistent by modifying the simple model so that host types are interpreted as directed complete partial orders (dcpo's).

\newcommand\blfootnote[1]{%
  \begingroup
  \renewcommand\thefootnote{}\footnote{#1}%
  \addtocounter{footnote}{-1}%
  \endgroup
}

Many quantum algorithms are actually parameterized in the number of qubits that they operate on.
 For example, the Quantum Fourier Transform (QFT)
has a uniform definition for any number of qubits by recursion, where $H$ is the Hadamard gate
$\frac{1}{\sqrt{2}}\left(\begin{smallmatrix}1 &1 \\ 1 &-1\end{smallmatrix}\right)$
and
$R_n$ is the $Z$ rotation gate $\left(\begin{smallmatrix}1 &0\\ 0 &e^{2\pi i/2^n} \end{smallmatrix}\right)$.

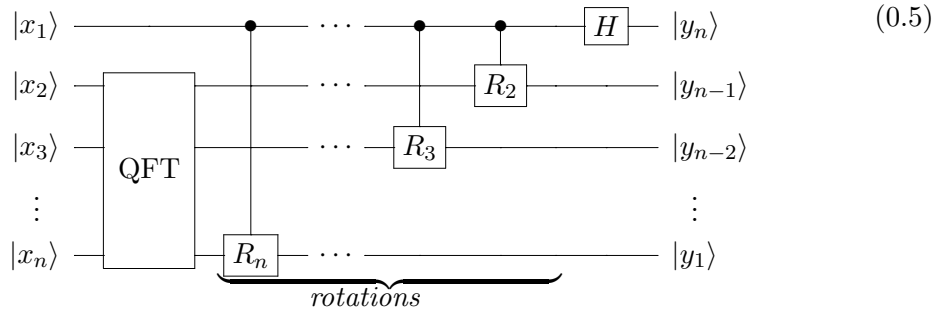
\begin{equation}
\centering
\scalebox{1}{     \Qcircuit @C=1em @R=.7em {
  \lstick{\ket{x_1}}      & \qw                       & \ctrl{4}     & \qw  & \cdots & & \ctrl{2}     &\ctrl{1}     &\qw & \gate{H} & \rstick{\ket{y_n}} \qw \\
  \lstick{\ket{x_2}}      & \multigate{3}{\text{QFT}} & \qw          & \qw  & \cdots & & \qw          & \gate{R_{2}}&\qw & \qw      & \rstick{\ket{y_{n-1}}} \qw \\
  \lstick{\ket{x_3}}      & \ghost{\text{QFT}}        & \qw          & \qw  & \cdots & & \gate{R_{3}} &  \qw        &\qw & \qw      & \rstick{\ket{y_{n-2}}} \qw \\
  \lstick{\vdots\ \ }     & \pureghost{\text{QFT}}    &              &      & & &                   &             &          & & \rstick{\ \ \vdots} \\
  \lstick{\ket{x_n}}      & \ghost{\text{QFT}}        & \gate{R_n}   & \qw  & \cdots & & \qw          &  \qw        & \qw      &\qw & \rstick{\ket{y_1}} \qw
\gategroup{1}{3}{5}{9}{4pt}{_\}}
\\&&&&&\textit{rotations}
 }}
\label{dgm:qft}
\end{equation}
\ \\
We formalize this by extending the circuit language with a wire type $\mathrm{QList}$ of qubit-lists for which the following equivalence of types holds:
\[
\mathrm{QList}\ \cong\
(\mathrm{qubit}\otimes \mathrm{QList}) \oplus 1
\]
so that we can define a function
\[
\mathrm{fourier}:\mathrm{Circ}(\mathrm{QList},\mathrm{QList})
\]

In practice, it will be useful for a circuit layout engine to know the number of qubits in the lists, suggesting a dependent type such as
\[\mathrm{fourier}:(n:\mathrm{Nat})\to \mathrm{Circ}(\mathrm{QList}(n),\mathrm{QList}(n))\]
but we leave these kinds of elaboration of the type system to future work. 

The categorical essence of recursive data types is algebraic compactness. In short, one says that a category $\cat{C}$ is algebraically compact (for a specific class of endofunctors) when every endofunctor $F:\cat{C}\to\cat{C}$ has a canonical fixpoint, which is the initial F-algebra~\cite{barr-algebraically}.
In earlier work~\cite{rennela-mfps30}, the first author has shown that the category of W*-algebras is algebraically compact and enriched in dcpo's, and so this is a natural
candidate for a semantics of the language (Section~\ref{sec:ewire-domains}).
In brief: circuit types are interpreted as W*-algebras,
and circuits are interpreted as completely positive sub-unital maps;
host types are interpreted as dcpo's;
in particular
the collection of circuits $\mathrm{Circ}(W,W')$ is interpreted as the dcpo of completely positive sub-unital maps, with the L\"owner order.
In this way, we provide a basic model for a quantum type theory with recursive types. 

\bigskip
\myparagraph{Relation to previous work}
This paper expands and develops the paper~\cite{mfps} presented at The Thirty-third Conference on the Mathematical Foundations of Programming Semantics (MFPS XXXIII). This version includes numerous elaborations, most notably Section~\ref{sec:LNL} on linear/non-linear models is entirely new.

\section{Functional programming and quantum circuits}%
\label{sec:qwire}

We introduce a new calculus called EWire
as a basis for analysing the basic ideas of embedding a circuit language inside a host functional programming language.
EWire (for `embedded wire language') is based on QWire~\cite{qwire} (`quantum wire language'),
and we make the connection precise in Section~\ref{sub:qwire}.  We discuss extensions of EWire in Section~\ref{sec:ewire-domains}.

We assume two classes of basic wire types.
\begin{itemize}
\item Classical wire types,
ranged over by $\cwA,\cwB,\dots$.
The wire types exist in both the circuit language and the host language.
For example, the type of classical bits, or Booleans.
\item Circuit-only wire types,
ranged over by $\qwA,\qwB,\dots$.
These wire types only exist in the circuit language.
For example, the type of qubits.
\end{itemize}
From these basic types we build all wire types:
\[
W,W' ::= I \mid W \otimes W'~|~\cwA~|~\cwB~|~\qwA~|~\qwB\dots
\]
We isolate the classical wire types, which are the types not using
any circuit-only basic types:
\[
V,V' \ ::=\ I~|~V\otimes V'~|~\cwA~|~\cwB\dots
\]
We also assume a collection $\mathcal{G}$ of basic gates, each assigned an input and an output wire type. We write $\mathcal{G}(W_{\text{in}},W_\text{out})$ for the collection of gates of input type $W_{\text{in}}$ and output type $W_{\text{out}}$.

In addition to the embedded circuit language, we consider a host language.
This is like Moggi's monadic metalanguage~\cite{moggi} but with special types
for the classical wire types $\cwA$, $\cwB$ and a type
$\Circ(W,W')$ of circuits for any wire types $W$ and $W'$. So the host types are
\[
A,B \ ::=\ A\times B~|~1~|~A\to B~|~\monad (A)~|~\Circ(W,W')~|~\cwA~|~\cwB
\]
The monad $T$ is primarily to allow probabilistic computations, although one might also add other side effects to the host language.
Notice that every classical wire type $V$
can be understood as a first order host type $\abs V$, according to the simple translation, called \textit{lifting}:
\[
\abs{V\otimes V'}\defeq \abs V\times \abs {V'}
\qquad
\abs{I} \defeq 1
\qquad
\abs{\cwA}\defeq \cwA
\]

\subsection{Circuit typing and host typing}
To describe the well-formed terms, we consider two kinds of context:
circuit contexts
\[\Omega = (w_1:W_1 \cdots w_n:W_n) \qquad \text{ (for $n \in \N$).}\]
and host language contexts
\[\Gamma = (x_1:A_1 \cdots x_m:A_m) \qquad \text{ (for $m \in \N$)}.\]
We will define a well-formed circuit judgement
\[\Gamma; \Omega \vdash C:W\]
and a well-formed host language judgement
\[\Gamma \vdash t:A\text.\]

Following QWire~\cite{qwire}, the circuit language is based on patterns.
Wires are organised in patterns given by the grammar
\[
p ::= w \mid () \mid (p,p)
\]
associated to the following set of rules, defining a ternary relation
$\Omega\implies p :W$:
\[
\begin{prooftree}
-\justifies
\cdot \implies ():1
\end{prooftree}
\qquad
\begin{prooftree}
-\justifies
w:W \implies w:W
\end{prooftree}
\]
\[
\begin{prooftree}
\Omega_1 \implies p_1:W_1 \qquad \Omega_2 \implies p_2:W_2
\justifies
\Omega_1, \Omega_2 \implies (p_1,p_2):W_1 \otimes W_2
\end{prooftree}
\qquad
\begin{prooftree}
\Omega_1,w_1:W_1,w_2:W_2,\Omega_2 \implies p:W
\justifies
\Omega_1,w_2:W_1,w_1:W_2,\Omega_2 \implies p:W
\end{prooftree}
\]

\myparagraph{Linear type theory for circuits}
The first five term formation rules are fairly standard for a linear type theory.
These are the constructions
for sequencing circuits, one after another, and ending by outputting
the wires, for splitting a tensor-product type into its constituents,
and the exchange rule.
The sixth rule includes the basic gates in the circuit language.
\[
\begin{prooftree}
\Gamma; \Omega_1 \vdash C_1:W_1 \qquad \Omega \implies p:W_1 \qquad \Gamma; \Omega, \Omega_2 \vdash C_2:W_2
\justifies
\Gamma; \Omega_1, \Omega_2 \vdash p \leftarrow C_1;C_2:W_2
\end{prooftree}
\qquad
\begin{prooftree}
\Omega \implies p:W
\justifies
\Gamma; \Omega \vdash \textbf{output } p:W
\end{prooftree}
\]
\[
\begin{prooftree}
\Omega \implies p:1 \quad \Gamma; \Omega'\vdash C:W
\justifies
\Gamma; \Omega, \Omega' \vdash () \leftarrow p; C:W
\end{prooftree}
\qquad
\begin{prooftree}
\Omega \implies p:W_1 \otimes W_2 \quad
\Gamma; w_1:W_1, w_2:W_2, \Omega' \vdash C:W
\justifies
\Gamma; \Omega,\Omega' \vdash (w_1,w_2) \leftarrow p; C:W
\end{prooftree}
\]
\[
  \begin{prooftree}
    \Gamma; \Omega_1,w_1:W_1,w_2:W_2,\Omega_2 \vdash C:W
    \justifies
    \Gamma; \Omega_1,w_2:W_2,w_1:W_1,\Omega_2 \vdash C:W
  \end{prooftree}
\]
\[
\begin{prooftree} \Omega_1 \implies p_1: W_1 \qquad \Omega_2 \implies p_2: W_2 \qquad
\Gamma;\Omega_2,\Omega\vdash C:W
\justifies
\Gamma; \Omega_1,\Omega \vdash p_2 \leftarrow \textbf{gate } g\, p_1; C:W
\using {g \in \mathcal{G}(W_1,W_2) }\end{prooftree}
\]

For example, qubit-based coin flipping is given by the following circuit:
\begin{equation}\label{eqn:flip}
\text{flip} \defeq a\leftarrow \textbf{gate}~\mathrm{init}_0~() ; a'\leftarrow \textbf{gate}~\mathrm{H}~a;b\leftarrow \textbf{gate}~\mathrm{meas}~ a' ; \textbf{output}~b
\end{equation}
In quantum circuit notation, this would be notated:
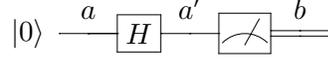
\[
\Qcircuit @C=1em @R=.7em {
\lstick {\ket 0} & \ustick a \qw & \gate{H}\qw & \ustick {a'}\qw & \meter&\ustick b\cw &\cw
}\]
\ \\
\myparagraph{Interaction between the circuits and the host}
A well-formed host judgement $\Gamma \vdash t: A$ describes a host-language program of type $A$ in the context of host language variables $\Gamma$.
The next set of typing rules describe the interaction between the
host language and the circuit language.
The host can run a circuit and get the result.
Since this may have side effects, for example probabilistic behaviour,
it returns a monadic type.
\[
\begin{prooftree}
\Gamma; \cdot \vdash C: V
\justifies
\Gamma \vdash \textbf{run } C:\monad(\abs V)
\using V \textit{ classical}
\end{prooftree}
 \]
 For example,
 $\text{flip}$~\eqref{eqn:flip} is an expression in the circuit language
 of type $\tbool$, and
 $(\textbf{run }\text{flip}):T(\tbool)$ is an expression of
 the host language that performs the coin flip.

The next two rules concern boxing a circuit as data in the host language, and then unboxing the data to form a
circuit fragment in the circuit language.
Notice that unboxing requires a pure program of type $\Circ(W_1,W_2)$,
rather than effectful program of type $T(\Circ(W_1,W_2))$.
For example, you cannot unbox a probabilistic combination of circuits.
The monadic notation clarifies this point.
\[\begin{array}{c}
\begin{prooftree}
\Omega \implies p:W_1 \qquad \Gamma;\Omega \vdash C:W_2
\justifies
\Gamma \vdash \textbf{box } (p:W_1) \Rightarrow C: \text{Circ}(W_1,W_2)
\end{prooftree}
\qquad\qquad
\begin{prooftree}
\Gamma \vdash t:\text{Circ}(W_1,W_2) \quad \Omega \implies p:W_1
\justifies
\Gamma; \Omega\vdash \textbf{unbox } t\, p:W_2
\end{prooftree}
  \end{array}\]
Finally we consider dynamic lifting, which, informally, allows us to send classical data
to and from the host program while the quantum circuit is running.
\[
\begin{prooftree}
\Gamma \vdash t:\abs V
\justifies
\Gamma;\cdot\vdash \textbf{init }t:V
\using V \textit{ classical}
\end{prooftree}
\qquad\qquad
\begin{prooftree}
\Omega \implies p: V \qquad \Gamma,x:\abs V;\Omega'\vdash C:W
\justifies
\Gamma; \Omega,\Omega' \vdash x \Leftarrow \textbf{lift } p; C:W
\using{V \textit{ classical}}
\end{prooftree}
\]
For example, we can define  a host function
\[(|V| \to \Circ(W_1,W_2))\to \Circ(V\otimes W_1,W_2)\]
converting a host language parameter to a circuit wire,
by
\[
  \lambda f.\,\textbf{box }\Big((v,w)\Rightarrow \Big(
  x\Leftarrow \textbf{lift }v; \textbf{unbox}(fx)(w)\Big)\Big)
\]
If $V=\tbool$, then this might be informally notated
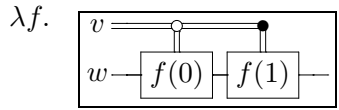
\[
\lambda f.\quad
\framebox{  \Qcircuit @C=.5em @R=.7em {
    v&&\cw&\controlo{1}\cw&\control{1}\cw\\
    w&&\qw&\gate{f(0)}\qw\cwx &\gate{f(1)}\cwx\qw &\qw&\qw
}}\]
This function has an inverse,
\[
\Circ(V\otimes W_1,W_2)\to   (V \to \Circ(W_1,W_2))
\]
transforming a classical circuit wire to a parameter in the host language:
\[
  \lambda C.\,\lambda x.\,\textbf{box }\Big(w\Rightarrow \Big(
  v\Leftarrow \textbf{init }x;\textbf{unbox }(C)(v,w)\Big)\Big)\text.
\]
This might be informally notated\[
\lambda C.\lambda x.\quad
\framebox{ \, 
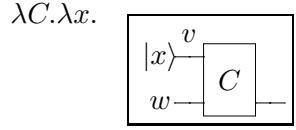
\Qcircuit @C=.5em @R=.7em {\\
    \ket x&&\ustick v\qw &\multigate{1}C&
    \\
    w&&\qw &\ghost C&\qw&\qw
}}\]
As we will see in (\ref{eqn:copower}), these functions together
correspond to a copower structure in enriched category theory.

For a further example of these interactions,
if we had conditionals in the host language, then
the illustration of classical control in the introduction~\eqref{qcircuit-example} could be notated
\begin{equation}\begin{array}{l@{}l}
    {  \cdot};a:\tqubit,b:\tqubit\vdash\

  &
    a'\leftarrow \text{meas }a;
    x\Leftarrow \textbf{lift }a';\\
  &\begin{array}{l@{}l}
    \textbf{unbox}\Big(&
    \text{if }x\text{ then }\textbf{box}(b'\Rightarrow (b''\leftarrow \textbf{gate }X \,b';\textbf{output }b''))\\
                       &\text{ else }\textbf{box}(b'\Rightarrow \textbf{output }b')\Big)\end{array}
                         \end{array}
\end{equation}
Notice that the classical control is handled in the host language.

\myparagraph{Additional typing rules for the host language}
Recall that the types of the host language are
\[A,B \ ::=\ A\times B~|~1~|~A\to B~|~\monad (A)~|~\Circ(W,W')~|~\cwA~|~\cwB\]
The standard typing rules of the monadic metalanguage are the rules of
the simply-typed $\lambda$-calculus
\[
\begin{prooftree}
(x: A) \in \Gamma
\justifies
\Gamma \vdash x:A
\end{prooftree}
\qquad
\begin{prooftree}
\Gamma, x:A\vdash t:B
\justifies
\Gamma\vdash (\lambda x^A. t):A \to B
\end{prooftree}
\qquad
\begin{prooftree}
\Gamma \vdash t:A \to B
\qquad
\Gamma \vdash u:A
\justifies
\Gamma \vdash t(u):B
\end{prooftree}
\]
Terms of product types are formed following four typing rules
\[
\begin{prooftree}
-\justifies
\Gamma \vdash \text{unit}:1
\end{prooftree}
\qquad
\begin{prooftree}
\Gamma \vdash t:A \quad \Gamma \vdash u:B
\justifies
\Gamma \vdash (t,u):A \times B
\end{prooftree}
\qquad
\begin{prooftree}
\Gamma \vdash t:A \times B
\justifies
\Gamma \vdash \pi_1(t):A
\end{prooftree}
\qquad
\begin{prooftree}
\Gamma \vdash t:A \times B
\justifies
\Gamma \vdash \pi_2(t):B
\end{prooftree}
\]
to which we need to add the typing rules for the monad~\cite{moggi}, associated respectively to the unit and the strong Kleisli
composition:

\[
\begin{prooftree}
\Gamma \vdash t:B
\justifies
\Gamma \vdash \ireturn t:T(B)
\end{prooftree}
\qquad
\begin{prooftree}
\Gamma \vdash t:T(B)
\qquad
\Gamma,x:B\vdash u : T(C)
\justifies
\Gamma \vdash \iletin t x u : T(C)
\end{prooftree}
\]
This is in addition to the typing rules for the interaction between the host language and the circuit language above.


\subsection{QWire}%
\label{sub:qwire}

The language QWire of Paykin, Rand and Zdancewic~\cite{qwire} is an instance of EWire where:
\begin{itemize}
\item there is one classical wire type, $\tbool$,
and one circuit-only wire type, $\tqubit$.
\item
there are basic gates $\meas\in \mathcal G(\tqubit,\tbool)$
and $\newq\in\mathcal G(\tbool,\tqubit)$.
\end{itemize}
A subtle difference between EWire and QWire is that in QWire one can directly run a circuit of type qubit, and it will produce a bit, automatically measuring the qubit that results from the circuit.
To run a circuit of type $\tqubit$ in EWire, one must append an explicit measurement at the end of the circuit.
These explicit measurements can be appended automatically,
to give a translation from QWire proper to this instantiation of EWire.

We now summarize how this is done.
We first define a translation $(\classicalize-)$ from
\emph{all} wire types to classical wire types:
\[\classicalize {W\otimes W'}\defeq \classicalize W\otimes \classicalize W'
\quad
\classicalize I\defeq I
\quad
\classicalize \tbool\defeq \tbool
\quad
\classicalize \tqubit\defeq \tbool
\]
Then, from an arbitrary wire type $W$, we can
extract a host type $\abs{\classicalize W}$.

From the basic gates $\meas$ and $\new$ we can define circuits
\[
\meas_W:\Circ(W,\classicalize W)
\qquad
\text{ and }
\qquad
\new_W:\Circ(\classicalize W,W)
\]
for all wires $W$.
These are defined by induction on $W$. For example,
\[
\meas_{I}\defeq \tid
\quad
\meas_{\tbool} \defeq \tid
\]
\[
\meas_{\tqubit} \defeq \tbox p \Rightarrow p' \leftarrow \tgate \meas p; \toutput p'
\]
\begin{align*}
\meas_{W \otimes W'}
\defeq
\tbox(w,w') \Rightarrow\ \
&x \leftarrow \tunbox \meas_W w;\ x' \leftarrow \tunbox \meas_{W'} w';\\
&\toutput (x,x')
\end{align*}
and $\new_W$ is defined using $\new \in \mathcal G(\tbool,\tqubit)$ similarly.
Then we define the following derived syntax,
so that run and lift can be used at all wire types, not just the classical ones:
\begin{align*}
&\mathbf{qwire\text-run}(C)\ \defeq \ \mathbf{run}(x\leftarrow C; \mathbf{unbox}\,\meas\,x)\\
&(x\Leftarrow \mathbf{qwire\text-lift}\,p\,;\,C)\ \defeq \
y\Leftarrow\mathbf{lift}\,p\,;\,x\leftarrow \mathbf{unbox}\,\new\,y\,;\,C
\end{align*}


\section{Categorical models of EWire}%
\label{sec:cat-qwire}
We introduce the categorical semantics of EWire. Our semantic models are based around enriched category theory~\cite{kelly}.
The relevance of $\mathbf{Set}$-enriched copowers to quantum algorithms has previously
been suggested by Jacobs~\cite{jacobs-block}. On the other hand, copowers and enrichment play a key role in the non-quantum
enriched effect calculus~\cite{enriched-effect-calculus,ms-state} and other areas~\cite{levy-CBPV-book,mellies-enriched-adjunctions,power-lawvere,staton-freyd-lawvere}. Nonetheless, the connection that we establish with the EWire syntax appears to be novel.

\subsection{Preliminaries}
\myparagraph{Enriched categories}
Let $\cat H$ be a cartesian closed category. Recall that a category $\cat C$ enriched in $\cat H$ is given by a collection of objects
and, for each pair of objects $c$ and $d$, an object $\cat C(c,d)$ in $\cat H$,
together with morphisms for composition $\cat C(c,d)\times \cat C(d,e)\to \cat C(c,e)$ and
identities $1\to \cat C(c,c)$, subject to associativity and identity laws.
Much of category theory can be reformulated in the enriched setting~\cite{kelly}.

\medskip

\myparagraph{Computational effects} Embedding the circuit language requires the use of some computational effects in the host language.
When the circuit language involves quantum measurement, then the closed host term
$
\vdash \textbf{run}\big(\text{flip}\big) : T(\tbool)
$
is a coin toss, and so the semantics of the host language must accommodate probabilistic features.

Following Moggi, we model this by considering a cartesian closed category $\cat H$ with an enriched monad on it.
Recall that an enriched monad is given by an endofunctor $T$ on $\cat H$ together with
a unit morphism $\eta : X\to T(X)$ for each $X$ in $\cat H$,
and a bind morphism
\[
\cat H(X,T(Y))\to \cat H(T(X), T(Y))
\]
for objects $X$ and $Y$, subject to the monad laws~\cite{moggi}.

The idea is that deterministic, pure programs in the host language are interpreted as morphisms in $\cat H$.
Probabilistic, effectful programs in the host language are interpreted as Kleisli morphisms, i.e.\ morphisms $X\to T(Y)$.

\medskip

\myparagraph{Relative monads}
The monads of Moggi provide a first class type of computations $T(A)$.
In a truly first order language, such as a circuit language, although computations are present, they are not first class. In particular,
there is no \emph{wire} type $T(\tqubit)$ of all quantum computations.
To resolve this kind of mismatch, authors have proposed alternatives such as relative monads~\cite{altenkirch-monads} and monads with arities~\cite{berger-mellies-weber}.

An ordinary \textit{relative adjunction} is given by three functors $J:\cat{B}\to\cat{D}$, $L:\cat{B}\to\cat{C}$ and $R:\cat{C}\to\cat{D}$ such that there is a natural bijection
\begin{equation}\label{eqn:rel-adj}
\cat{C}(L(b),c)\cong\cat{D}(J(b),R(c))
\end{equation}
We write $L {\,\,}_J\!\dashv R$ and call \textit{relative monad} the functor $RL:\cat{B}\to\cat{D}$.

Enriched relative adjunctions and enriched relative monads are defined in the obvious way, by requiring $J$, $L$ and $R$ to be enriched functors and replacing the natural bijection~\eqref{eqn:rel-adj} with a antural isomorphism.
In an enriched relative monad $T=RL$, the bind operation is a morphism of type
\[
\cat D(J(X),T(Y))\to \cat D(T(X), T(Y))
\]

\myparagraph{Copowers and weighted limits}
In enriched category theory it is appropriate to consider weighted colimits and limits, which are a generalization of the ordinary notions.
As a first step, we consider copowers. A copower is a generalization of an $n$-fold coproduct.

Let $n$ be a natural number, and
let $A$ be an object of a category $\cat C$ with sums. The copower $n\odot A$
is the $n$ fold coproduct $A + \cdots + A$.
This has the universal property that to give a morphism $n\odot A\to B$
is to give a family of $n$ morphisms $A\to B$.
In general, if $\cat C$ is a category enriched in a category $\cat H$,
and $A$ is an object of $\cat C$ and $h$ an object of $\cat H$, then
the \emph{copower} is an object $h\odot A$ together with a family of isomorphisms
\begin{equation}\label{eqn:copower}
\cat C(h\odot A,B)\cong \cat H(h,\cat C(A,B))
\end{equation}
natural in $B$.

Weighted colimits combine the notions of copowers and of ordinary conical colimits.
If $\cat J$ is a $\cat H$-enriched category, then we may try to take a colimit of a diagram $D:\cat J\to \cat C$
weighted by a functor $W:\cat J\op\to\cat H$. Here $\cat C$ is an $\cat H$-enriched category and $D$ and $W$ are $\cat H$-functors.
A \emph{cylinder} for $(W,D)$ is given by an object $X$ and a $\cat H$-natural family of morphisms $W(j)\to \cat C(D(j),X)$;
the weighted limit, $\colim^W_j D_j$ if it exists, is the universal cylinder.

\subsection{EWire models}
Let us define a sufficient set of properties which ensure that a pair of categories corresponds to a categorical model in which one can interpret EWire, in order to reason about circuits and identify their denotational meaning. We assume that the circuit language is parametrized by a fixed collection of gates, noted $\mathcal{G}$.

\begin{defi}%
\label{def:model-QWire}

A \emph{categorical model of EWire} $(\cat C,\cat H,\cat H_0,T)$ is given by the following data:
\begin{enumerate}
\item A cartesian closed category $\cat H$ with
  a strong monad $T$ on $\cat H$. This is needed to interpret the host language.
\item A small full subcategory $j:\cat H_0\subseteq \cat H$.
  The idea is that the objects of $\cat H_0$ interpret the first order host types, equivalently, the classical wire types: the
  types that exist in both the host language and the circuit language.
\item\label{enum:interp-circuits} An $\cat H$-enriched symmetric monoidal category $(\cat{C},\otimes,I)$.
  This allows us to interpret the circuit language, and the $\cat H$-enrichment
  allows us to understand the host types $\Circ(W,W')$.
\item\label{enum:copow}  The category $\cat C$ has copowers by the objects of $\cat{H}_0$.
 The copower induces a functor $J:\cat{H}_0 \to \cat{C}$ defined by $J(h) = h \odot I$. Then, we have a natural isomorphism
 \[
 \cat{C}(J(h), C)=\cat{C}(h \odot I, C) \cong \cat{H}(j(h),\cat{C}(I,C))
 \]
  and therefore a $j$-relative adjunction \hide{$\xymatrix{\cat{C} & \cat{H}_0 \ltwocell_{\cat{C}(I,-)}^J{'\top_J} }$} $J {\,\,}_j\!\!\dashv \cat{C}(I,-)$ between circuits and (host) terms.
This functor $J:\cat H_0\to \cat C$ interprets the translation between first order host types and classical wire types.
 \item\label{enum:SMC-copow} For each object $A$ of $\cat C$, the functor $A\otimes -:\cat C\to \cat C$ preserves copowers.
This makes the functor $J$ symmetric monoidal,
and makes the relative adjunction an enriched relative adjunction.
 \item\label{enum:monad} There is an enriched relative monad morphism
 \[
 \text{run}_h:\cat{C}(I,J(h)) \to T(j(h))
 \]
 where the enriched relative monad $\cat{C}(I,J(-)):\cat{H}_0\to\cat{H}$ is induced by the enriched $j$-relative adjunction
${J {\,\,}_j\!\!\dashv \cat{C}(I,-)}$.
\hide{whose unit and bind are respectively given by the morphisms:
\[n \cong \cat{H}_0(1,n) \to \cat{C}(J(1),J(n)) \cong \cat{C}(I,J(n))\]
\[\cat{H}(m,\cat{C}(I,J(n))) \times \cat{C}(I,J(m)) \cong \cat{C}(J(m),J(n)) \times \cat{C}(I,J(m)) \to \cat{C}(I,J(n))\]}
This is the interpretation of running a quantum circuit, producing some classical probabilistic outcome.
\end{enumerate}
\end{defi}

\noindent
If the category $\cat C$ has a given
object $\denot \qwA$ for each basic quantum wire type $\qwA$,
and $\cat H_0$ has a given object
$\denot \cwA$ for each basic classical wire type $\cwA$,
then we can interpret all wire types $W$ as objects of $\cat C$:
\[\denot 1 \defeq I
\quad
\denot {\cwA}\defeq  J(\denot{\cwA})
\quad
\denot{W\otimes W'}\defeq \denot W\otimes \denot{W'}\text.
\]
If the category $\cat C$ also has a given morphism
$\denot{g}:\denot {W_1}\to\denot{W_2}$ for every gate $g \in \mathcal{G}(W_1,W_2)$,
then we can interpret the circuit langauge inside $\cat C$.

In light of those axioms, and to every categorical model of EWire, we associate the following denotational semantics. \hide{We refer the interested reader to~\cite{qwire}[App.~B] for a proof of soundness.}

First, we define as promised the denotation of the host type $\text{Circ}(W,W')$ by
\[
\denot{\text{Circ}(W,W')} \defeq \cat{C}(W,W') \in \text{Obj}(\cat{H})
\]
The semantics of the other host types is given as follows:
\[\denot {1} \defeq 1
\quad
\denot{A\times A'}\defeq \denot A\times \denot{A'}
\quad
\denot{A\to A'}\defeq (\denot A\to \denot{A'})
\quad
\denot{T(A)}\defeq T(\denot A)\text.
\]
Ordered context of wires $\Omega$ have the following semantics:
\[
\denot{\langle \cdot \rangle}=I
\qquad
\denot{w:W}=\denot{W}
\qquad \denot{\Omega, \Omega'} = \denot{\Omega} \otimes \denot{\Omega'}
\]
A circuit judgement $\Gamma;\Omega\vdash t:W$ is denoted by \[
\denot{\Gamma;\Omega\vdash t:W} \in \cat{H}(\denot{\Gamma},\cat{C}(\denot{\Omega},\denot{W}))
\]
relying on the assumption that the category $\cat{H}$ is a model of the host language. A host type $\text{Circ}(W,W')$ is interpreted as the hom-object $\cat{C}(\denot{W},\denot{W'})$, in the category $\cat{H}$. In this setting, denotations of boxing and unboxing instructions are trivial. Indeed, notice that whenever $\Omega \implies p:W_1$ holds, we have $\denot{\Omega}\cong\denot{W_1}$, and we put
\[\denot{\Gamma \vdash \textbf{box } (p:W_1) \Rightarrow C:\text{Circ}(W_1,W_2)}=\denot{\Gamma;\Omega\vdash C:W_2}
\]
\[
\denot{\Gamma;\Omega\vdash \textbf{unbox } t\, p:W_2}=\denot{\Gamma \vdash t: \text{Circ}(W_1,W_2)}
\]


The denotation of $\textbf{output } p:W$ is the identity.
Moreover, instructions $\Gamma; \Omega, \Omega' \vdash () \leftarrow p; C:W$ and $\Gamma; \Omega' \vdash C:W$ (resp. $\Gamma; \Omega, \Omega' \vdash (w_1,w_2) \leftarrow p; C:W$ and $\Gamma; w_1:W_1, w_2:W_2, \Omega' \vdash C:W$) have isomorphic denotations whenever $\Omega \implies p:1$ holds (resp. $\Omega \implies p: W_1 \otimes W_2$ holds).

The \textbf{lift} construction is interpreted by the copower. In detail, for every object $h$ of $\cat{H}$, and every object $h'$ of $\cat H_0$, we consider the isomorphism
\begin{align*}
\text{lift}_{h'}: \cat{H}(h\times h',\cat{C}(X, Y)) \cong \cat{H}(h,\cat{H}(h',\cat{C}(X,Y))) & \cong \cat{H}(h, \cat{C}(h'\odot X,Y))\\ & \cong \cat H(h,\cat C(J(h')\otimes X,Y))
\end{align*}
so that
\[
\denot{\Gamma; \Omega,\Omega' \vdash x \Leftarrow \textbf{lift } p; C:W}
= \text{lift}_{\denot{\abs V}}(\denot{\Gamma,x:\abs V;\Omega'\vdash C:W})
\]

As typing rules for the monad, the operations \textbf{return} and \textbf{let} are denoted respectively by the unit and the strong Kleisli composition~\cite{moggi}.

Since we're enforcing explicit measurement here, the denotation of the operation \textbf{init} for a term $t$ of first order host type $|V|$ (for a classical wire type $V$) and the denotation of the operation \textbf{run} for a circuit $C$ whose output wire type is the type $W$ are given by Def.~\ref{def:model-QWire}(\ref{enum:monad}). 
\[
\denot{\Gamma; \cdot\vdash \textbf{init }t:V} =
 \denot \Gamma \xrightarrow{t} \denot{V} \xrightarrow{\cat C(I,J(-))} \cat C(I,J(\denot{V}))
\]
\[
\denot{\Gamma \vdash \textbf{run }C:T(|V|)} =
 \denot \Gamma \xrightarrow{C} \cat C(I,J(\denot{V})) \xrightarrow{\text{run}_{\denot{V}}} T(\denot{V})
\]
The denotations of the remaining instructions are given by the following composite morphisms
in $\cat H$. They crucially use the enriched composition, which is a morphism in $\cat H$.
\begin{multline*}
\denot{\Gamma; \Omega_1,\Omega_2 \vdash p \leftarrow C; C': W'}= \\
  \denot \Gamma \xrightarrow {C,C'}\cat C(\denot{ \Omega_1},\denot W)\times \cat C(\denot{\Omega,\Omega_2},\denot{W'})
\   \stackrel p  \cong \
  \cat C(\denot{ \Omega_1},\denot{\Omega_2})\times \cat C(\denot{\Omega,\Omega_2},\denot{W'})
  \\\xrightarrow{(\text{inj}_\Omega\times -)}
  \cat C(\denot{ \Omega,\Omega_1},\denot{\Omega,\Omega_2})\times \cat C(\denot{\Omega,\Omega_2},\denot{W'})
  \xrightarrow \circ
  \cat C(\denot{\Omega,\Omega_1},\denot{W'})
\end{multline*}
\begin{multline*}
  \denot{\Gamma; \Omega_1,\Omega \vdash p_2 \leftarrow \textbf{gate } g\, p_1; C: W}=
  \\
  \begin{aligned}  \denot \Gamma \cong 1\times \denot \Gamma \xrightarrow {\denot g \times C}\ &
    \cat C(\denot {W_1},\denot{W_2}) \times \cat C(\denot{ \Omega_2,\Omega},\denot W)
    \\
  \stackrel {p_1,p_2}  \cong \ &
  \cat C(\denot{ \Omega_1},\denot{\Omega_2})\times \cat C(\denot{\Omega,\Omega_2},\denot{W'})
  \\\xrightarrow{(\text{inj}_\Omega \times -)}\ &
  \cat C(\denot{ \Omega,\Omega_1},\denot{\Omega,\Omega_2})\times \cat C(\denot{\Omega,\Omega_2},\denot{W'})
  \xrightarrow \circ
  \cat C(\denot{\Omega,\Omega_1},\denot{W'})
  \end{aligned}
  \end{multline*}

where $C,C'$ is the product map, and $\text{inj}_\Omega$ is the injection map which sends circuits in $\cat C(\denot{ \Omega_1},\denot{\Omega_2})$ to circuits in $\cat C(\denot{ \Omega,\Omega_1},\denot{\Omega,\Omega_2})$. The later corresponds to the operation of adding a wire to a circuit.

\subsection{Example model: finite dimensional C*-algebras enriched in sets}
Our view on the semantics of quantum computing relies on the theory of C*-algebras.
The positive elements of C*-algebras correspond to observables in quantum theory, and we understand quantum computations as linear maps that preserve positive elements, in other words, `observable transformers'.
For example, the observables of a qubit are positive elements of the non-commutative algebra $M_2$ of $2\times 2$ complex matrices, and the observables of a bit are positive elements of the commutative algebra $\CC^2$ of pairs of complex numbers.
Circuit expressions $(\cdot;(x:W)\vdash C:W')$ will be interpreted as completely positive unital
maps $\denot {W'}\to\denot W$.
The reverse direction is in common with predicate
transformer semantics for conventional programming.

In short, a \textit{(unital) C*-algebra} (e.g.~\cite{sakai})
is a vector space over the field of complex numbers
that also has multiplication, a unit and an involution,
satisfying associativity and unit laws for multiplication,
involution laws (e.g. $x^{**}=x$, ${(xy)}^*=y^*x^*$, ${(\alpha x)}^*=\bar \alpha (x^*)$) and
such that the spectral radius provides a norm making it a Banach space.

There are two crucial constructions of C*-algebras: matrix algebras and direct sums.
\textit{Matrix algebras} provide a crucial example of C*-algebras.
For example, as already mentioned, the algebra $\mat 2$ of $2\times 2$ complex matrices represents the type of qubits.
The \emph{direct sum} of two C*-algebras, $A\oplus B$, is the set of pairs with componentwise algebra
structure.
For instance, $\CC\oplus \CC$ represents the type of classical bits.
Every finite-dimensional C*-algebra is a direct sum of matrix algebras.

The tensor product $\otimes$ of finite dimensional C*-algebras is
uniquely determined by two properties: (i)~that $M_k\otimes M_l\cong M_{k\times l}$, and
(ii)~that $A\otimes (-)$ and $(-)\otimes B$ preserve direct sums.
In particular $M_k\otimes A$ is isomorphic to the C*-algebra $M_k(A)$ of $(k\times k)$-matrices valued in $A$.

We do not focus here on linear maps that preserve all of the C*-algebra structure,
but rather on completely positive maps.
An element $x \in A$
is \emph{positive} if it can be written in the form $x=y^* y$ for $y \in
A$. These elements correspond to quantum observables.
A map $f : A \rightarrow B$, linear between the underlying vector spaces, is \textit{positive} if it preserves positive elements. A linear map is \emph{unital} if it preserves the multiplicative unit. A linear map $f$ is \textit{completely positive} if the map
\[
(M_k\otimes f):M_k\otimes A \to M_k\otimes B
\]
is positive for every $k$.
This enables us to define a functor $C\otimes (-)$ for every finite dimensional C*-algebra $C$.
Thus finite dimensional C*-algebras and completely positive unital linear maps form a symmetric monoidal category.
There are completely positive unital maps corresponding to initializing
quantum data, performing unitary rotations, and measurement,
and in fact all completely positive unital maps arise in this way~(e.g.~\cite{selinger-qpl,staton-popl15}).

\begin{prop}%
\label{prop:B2:fdcpu-model}
The quadruplet $(\FdCPU^{\mathrm{op}},\cat{Set},\N,\DM)$ is a model of EWire, formed by the opposite category of the category {$\FdCPU$} of finite-dimensional C*-algebras and completely positive unital maps, the cartesian closed category $\cat{Set}$ of sets and functions, the skeleton {$\N$} of the category of finite sets and functions (which considers natural numbers as its objects), and the probability distribution monad $\mathcal D$ over $\cat{Set}$. In fact it is a model of QWire, with $\denot {\text{qubit}}\defeq M_2$ and $\denot{\text{bit}}\defeq \C\oplus \C$.
\end{prop}

\begin{proof}
The category $\cat{Set}$ of sets and functions is the canonical example of cartesian closed category, and the distribution monad $\mathcal{D}:\cat{Set}\to\cat{Set}$ is a strong monad.

The category $\opp\FdCPU$ of the opposite category of finite-dimensional C*-algebras and completely positive unital maps has a monoidal structure given by the tensor product of C*-algebras, finite sums given by direct sums and the C*-algebra $\C$ of complex numbers is the unit $I$.

The copower {$n \odot A$} of a natural number $n \in \N$ and a C*-algebra $A$ is the C*-algebra $n \odot A$, defined as the $n$-fold direct sum $A \oplus \cdots \oplus A$ like in~\cite{jacobs-block}.  We observe that the copower distributes over the coproduct,
i.e.
\[
n \odot (A \oplus B) = (n \odot A) \oplus (n \odot B)
\]
and that composition is multiplication, i.e.
\[
n \odot (m \odot A) = (nm) \odot B
\]

The copower $n\odot\CC$ is the C*-algebra $\CC^n$. Copowers are preserved by endofunctors $A \otimes -$.
\[
 A \otimes (n \odot B) = (A \otimes B) \oplus \cdots \oplus (A \otimes B) = n \odot (A \otimes B)
\]
We still need to verify that we have a relative adjunction. Observing that
\[
  \FdCPU(A, \CC^n) \cong \FdCPU(A,\CC) \times \cdots \times \FdCPU(A,\CC)
\]
one deduces that
\[
  \FdCPU(A, \CC^n) \cong \cat{Set}(n,\FdCPU(A,\CC))
\]
and therefore
\[
  \opp\FdCPU(\CC^n,A) \cong \cat{Set}(n,\opp\FdCPU(\CC,A))
\]
Then, the symmetric monoidal functor $J:\N\to\opp\FdCPU$ associates every natural number $n \in \N$ to the C*-algebra $\C^n$. The morphisms $\text{run}_n$ are given by the isomorphism
\[
\opp\FdCPU(\C,\C^n) =\FdCPU(\C^n,\C)\cong\DM(n)
\qquad
\text{\cite[Lemma 4.1]{furber-jacobs}}
\]
between states on $\C^n$ and the $n$-simplex
\[
\DM(n):=\{x \in \unit^n \mid \textstyle\sum_{i=1}^n x_i = 1\}
\qquad \qquad
\]

The semantics of types and gates is rather standard.
Probabilities are in particular complex numbers $\CC$ and a (classical) bit is therefore an element of the C*-algebra $\CC \oplus \CC$. \hide{The truth and falsum values correspond respectively to the pairs $(1,0)$ and $(0,1)$.} Moreover, $n$-qubit systems are modelled in the C*-algebra $M_{2^n}$.  In other words,
\[
\denot{\text{1}}=\CC
\qquad
\denot{\text{bit}}={\CC\oplus\CC}
\qquad
\denot{\text{qubit}}={M_2}
\qquad
\denot{{u}}={u^\dagger (-) u}
\text{ (for every unitary $u \in \mathcal{U}$)}
\]
\[
\text{meas}:\C\oplus\C\to M_2:(a,b)\mapsto\left(\begin{smallmatrix} a & 0\\ 0 & b\end{smallmatrix}\right)
\qquad
\text{new}:M_2 \to \C \oplus \C:\left(\begin{smallmatrix}
a & b\\
c & d
\end{smallmatrix}\right)\mapsto (a,b)
\]
and so on.
\end{proof}

\subsection{Soundness of normalization steps}%
\label{sub:B2:operational}
In~\cite{qwire}, a partial normalization procedure is proposed
for QWire expressions,
which is shown to be sound with respect to a semantics in completely positive maps~\cite[App.~A--B]{qwire}.
We now remark that the program equations that are used to justify the steps in the normalization procedure are sound in any EWire model.

We say that an equation between circuit expressions
\[\Gamma;\Omega\vdash C_1\equiv C_2:W
\]
is \emph{sound} if the interpretations are equal,
$\denot {C_1}=\denot{C_2}$,  in every EWire model.
\begin{prop}%
\label{prop:soundness}
  The following equations between circuits are sound:
  \begin{itemize}
    \item $\textbf{unbox } (\textbf{box } w \Rightarrow C)\, p \equiv C[w \mapsto p]$
    \item $p \leftarrow \textbf{output } p'; C \equiv C[p \mapsto p']$
\item $
w \leftarrow (p_2 \leftarrow \textbf{gate } g\, p_1;N);C
\equiv
p_2 \leftarrow \textbf{gate } g\, p_1; w \leftarrow N; C
$
\item $
w \leftarrow (x \Leftarrow \textbf{lift } p; C');C
\equiv
x \Leftarrow \textbf{lift } p; w \leftarrow C'; C$
\item $
() \leftarrow (); C \equiv C
$
\item $(w_1,w_2) \leftarrow (p_1,p_2); C \equiv C[\begin{smallmatrix}
 w_1 &\mapsto p_1\\ w_2 &\mapsto p_2
\end{smallmatrix}]
$
\item
$
w \leftarrow (() \leftarrow w'; N); C
\equiv
() \leftarrow w'; w \leftarrow N; C
$
\item
  $w \leftarrow ((w_1,w_2) \leftarrow w'; N); C
\equiv
(w_1,w_2) \leftarrow w'; w \leftarrow N; C
$
\end{itemize}
\end{prop}
\begin{proof}[Proof notes]
  These equations are all straightforward consequences of the interpretation in monoidal categories, except for the first one, which is different in character, but trivial.
\end{proof}
We remark that the categorical model suggests other equations that could be used for a more advanced normalization algorithm, such as the following equations
that come from the definition of copower~(\ref{eqn:copower})
\[
  x\Leftarrow \textbf{lift }p;\textbf{init }x
  \equiv
  p
  \qquad
  p\leftarrow \textbf{init }t;x\Leftarrow \textbf{lift }p;C
  \equiv
  C[x\mapsto t]\text.
\]

\section{Towards a model with recursive quantum datatypes: enrichment in dcpos}%
\label{sec:ewire-domains}

Recall that a dcpo is a partial order which has all directed joins. A monotone function between dcpo's is continuous if it preserves directed joins. Let $\cat{Dcpo}$ be the category of dcpo's and continuous functions (e.g.~\cite{abramsky-jung}).
The category is cartesian closed, hence a suitable situation
for modelling a host language, but also supports recursion.
If $D$ is a dcpo with a bottom element $\bot$,  then there is a fixed point combinator
\[
  Y:(D\to D)\to D
\]
given by $Y(f)=\bigvee_{i=1}^\infty f^n(\bot)$,
with the property that $f(Y(f))=Y(f)$.

Any ordinary category is trivially enriched in dcpo's, by considering the hom-sets as flat partial orders.
This leads us to an EWire model
\begin{equation}\label{eqn:cpudcpo}
(\FdCPU^{\mathrm{op}},\cat{Dcpo},\N,\mathcal V)
\end{equation}
where $\N$ is the subcategory of $\cat{Dcpo}$ comprising the flat, finite orders,
and $\mathcal V$ is a probabilistic powerdomain monad on dcpos (e.g.~\cite{jones-plotkin-powerdomain}).
We could use this model to interpret an extension of EWire with
recursion in the host language.

We can also model iteration in the circuit language.
One way to model iteration in the circuit language is to work with subunital maps. Recall that a map~$f$ between C*-algebras is subunital
if $f(1)\leq 1$, i.e.~$1-f(1)$ is positive (e.g.~\cite[\S 6]{selinger-qpl}). The subunital maps can be ordered by the L\"owner partial order: $f\leq g$ if and only if $(g-f)$ is a positive map.
This turns out to be a directed complete partial order between
finite dimensional C*-algebras, and so the category
$\Fd\CPSU$ is enriched in $\cat{Dcpo}$ in a more interesting way~\cite{rennela-mfps30,rennela-msc-thesis,selinger-qpl}.

Thus we have an EWire model:
\begin{equation}\label{eqn:cpsudcpo}
  (\Fd\CPSU^{\mathrm{op}},\cat{Dcpo},\N,\mathcal V)
\end{equation}

In this model, the runnable circuits correspond to subprobability distributions:
\[\Fd\CPSU(\mathbb C^n,\mathbb C)
  \cong
  \mathcal V(n)
  \cong
  \{x\in\unit^n~|~\textstyle\sum_{i=1}^n x_i\leq 1\}
\]
with, intuitively, $1-\sum_{i=1}^n x_i$ the probability of diverging.
In this model, the space of circuits $\denot{\Circ(W_1,W_2)}=\Fd\CPSU(\denot {W_2},\denot{W_1})$ is a dcpo with a bottom element (the zero map).

For this reason we can add to the
host language a family of fixed point combinators:
\[
  Y_{A,W_1,W_2}:\Big((A\to \Circ(W_1,W_2))\to (A\to \Circ(W_1,W_2)\Big)\to A \to \Circ(W_1,W_2)
\]

We can use this to recursively define circuits.
For a simple example, suppose we also add basic arithmetic functions to the host langauge, with standard interpretation in dcpos. Then the following function defines a family of circuits where \lstinline|Hs n| is the composition of $n$ Hadamard gates.
\begin{lstlisting}
Hs : int -> Circ(qubit,qubit)
Hs = Y(lambda Hs. lambda n.
       if n=0 then box q => output q
       else box q => q' <- gate H q ; unbox (Hs (n-1)) q')
\end{lstlisting}
Note that \lstinline|Hs (-1)| diverges, in the sense that it is interpreted by the zero map.

\medskip
\myparagraph{A model with recursive quantum data}
A still more elaborate model begins from the observation that the category
of W*-algebras and normal subunital maps is also enriched in $\cat{Dcpo}$ via
the L\"owner order.

This category $\opp\wCPSU$ is symmetric monoidal when equipped with the spatial tensor product~\cite{kornell}.
In earlier work~\cite{rennela-mfps30}, the first author has shown that
$\wCPSU$ supports the construction of recursive types: it is algebraically compact. Thus this is an EWire model that allows wire types encoding streams of qubits, lists of qubits, and so on.

Much like in~\cite[Section~5.6]{jacobs-recipe}, we use the restricted version of the monad of subvaluations $V'=\cat{dcGEMod}(\unit^{(-)},\unit)$ on $\Dcpo$, where the category $\cat{dcGEMod}$ is the category of directed-complete generalized effect modules and Scott-continuous effect module homomorphisms, also considered as a category of quantum predicates in~\cite{rennela-msc-thesis,rennela-staton-mfps31}.
Thus
\[(\wCPSU^{\mathrm{op}},\cat{Dcpo},\N,\mathcal V')
\]
is an EWire model.

In this model, we can interpret a wire type
\lstinline|qlist| with gates
\begin{itemize}
\item \lstinline|isempty| $\in \mathcal G(\mbox{\lstinline|qlist|},\mbox{\lstinline|bit|}\otimes\mbox{\lstinline|qlist|})$
\item \lstinline|headtail| $\in \mathcal G(\mbox{\lstinline|qlist|},\mbox{\lstinline|qbit|}\otimes\mbox{\lstinline|qlist|})$
\item \lstinline|nil| $\in \mathcal G(I,\mbox{\lstinline|qlist|})$
\item \lstinline|cons| $\in \mathcal G(\mbox{\lstinline|qbit|}\otimes\mbox{\lstinline|qlist|},\mbox{\lstinline|qlist|})$.
\end{itemize}
and a classical wire type \lstinline|int|.

We want to implement the Quantum Fourier Transform (see~Figure~\ref{qft}). Taking inspiration from~\cite[Sec.~6.2]{qwire} and~\cite{quipper}, we assume a host language constant
\[
\textbf{CR}:\tnat \to \text{Circ}(\text{qubit}\otimes\text{qubit},\text{qubit}\otimes\text{qubit})
\]
so that
$(\textbf{CR}\,n)$ corresponds to the controlled rotation by $\frac{2\pi}{2^n}$ around the $z$-axis. Then the program \lstinline|rotations| performs the rotations of a QFT circuit and the instruction \lstinline|fourier| corresponds to the QFT, as illustrated in the circuit in the introduction.

\begin{figure}
\caption{\label{qft}Implementation of QFT}
\noindent\begin{minipage}{\textwidth}
\begin{lstlisting}
length : Circ(qlist,int(*$\otimes$*)qlist) =
box qs =>
 (b,qs) <- isempty qs ; b <- lift b ;
 unbox (if b then box qs => n <- init 0 ; output (n,qs)
        else box qs => (h,t) <- gate headtail qs ;
                       (n,t) <- unbox length t ;
                       n <- lift n ; n' <- init (n+1) ;
                       qs <- gate cons (h,t) ;
                       output (n',qs)            ) qs
\end{lstlisting}

\end{minipage}
\hfill
\noindent \begin{minipage}{\textwidth}
\begin{lstlisting}
rotations : int -> Circ(qubit(*$\otimes$*)qlist,qubit(*$\otimes$*)qlist) =
lambda m. box (c,qs) =>
 (b,qs) <- isempty qs ; b <= lift b ;
 unbox (if b then box (c,qs) => output(c,qs)
        else box (c,qs) =>
             (n,qs) <- unbox length qs ; n <= lift n ;
             (q,qs') <- gate headtail qs
             (c,qs') <- unbox (rotations m) (c,qs')
             (c,q) <- unbox (CR (m-n)) (c,q) ;
             qs <- gate cons (q,qs')
             output (c,qs) )
       (c,qs)
\end{lstlisting}
\end{minipage}
\hfill
\begin{minipage}{\textwidth}
\begin{lstlisting}
fourier : Circ(qlist, qlist) =
box qs =>
  (b,qs) <- isempty qs ; b <= lift b ;
  unbox (if b then box qs => output qs
         else box qs =>
              (q,qs') <- gate headtail qs;
              qs' <- unbox fourier qs';
              (n,qs') <- unbox length qs' ; n <= lift n;
              (q,qs') <- unbox (rotations n) (q,qs')
              q <- gate H q ; output (q,qs')    ) qs
\end{lstlisting}
\end{minipage}
\end{figure}

Here we are using some standard syntactic sugar for the host language, such as recursive definitions instead of a fixed point combinator.
The recursive definitions encode the following informal recursive circuit definitions:
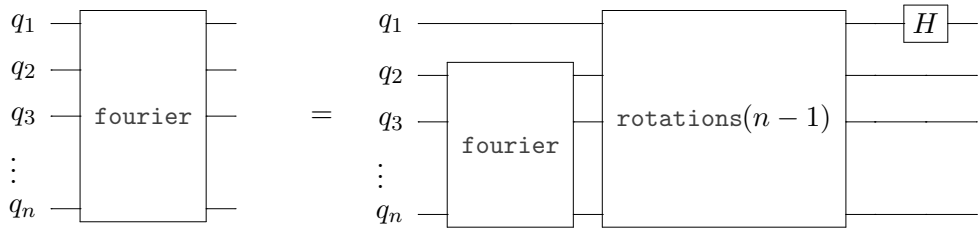
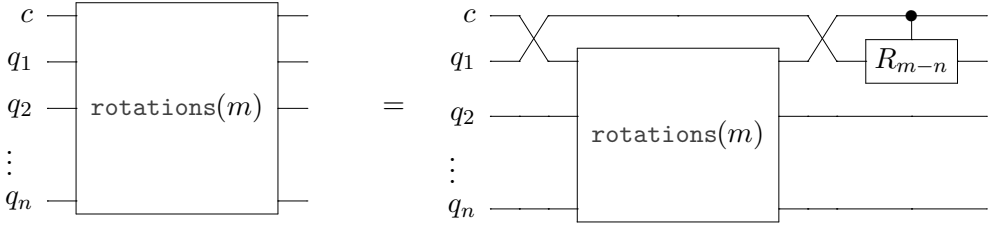
\[
\centering
\scalebox{1}{     \Qcircuit @C=1em @R=.7em {
  \lstick{q_1}      & \multigate{4}{\text{\lstinline|fourier|}} & \qw \\
  \lstick{q_2}      & \ghost{\text{\lstinline|fourier|}}        & \qw \\
  \lstick{q_3}      & \ghost{\text{\lstinline|fourier|}}        & \qw & & & = & & &\\
  \lstick{\vdots\ \ }& \pureghost{\text{\lstinline|fourier|}}   & \\
  \lstick{q_n}      & \ghost{\text{\lstinline|fourier|}}        & \qw
}}
\scalebox{1}{     \Qcircuit @C=1em @R=.7em {
  \lstick{q_1}      & \qw                                       & \multigate{4}{\text{\lstinline|rotations |}(n-1)} &\qw & \gate{H} & \qw \\
  \lstick{q_2}      & \multigate{3}{\text{\lstinline|fourier|}} & \ghost{\text{\lstinline|rotations |}(n-1)}        &\qw & \qw & \qw\\
  \lstick{q_3}      & \ghost{\text{\lstinline|fourier|}}        & \ghost{\text{\lstinline|rotations |}(n-1)}        &\qw & \qw & \qw\\
  \lstick{\vdots\ \ }& \pureghost{\text{\lstinline|fourier|}}   & \pureghost{\text{\lstinline|rotations |}(n-1)}    &    &     &    \\
  \lstick{q_n}      & \ghost{\text{\lstinline|fourier|}}        & \ghost{\text{\lstinline|rotations |}(n-1)}        &\qw & \qw & \qw\\
 }}
\]
\[
\centering
\scalebox{1}{     \Qcircuit @C=1em @R=.7em {
  \lstick{c}        & \multigate{4}{\text{\lstinline|rotations |}(m)} & \qw \\
  \lstick{q_1}      & \ghost{\text{\lstinline|rotations |}(m)}      & \qw \\
  \lstick{q_2}      & \ghost{\text{\lstinline|rotations |}(m)}      & \qw & & & = & & &\\
  \lstick{\vdots\ \ }& \pureghost{\text{\lstinline|rotations |}(m)} &  \\
  \lstick{q_n}      & \ghost{\text{\lstinline|rotations |}(m)}       & \qw
}}
\scalebox{1}{     \Qcircuit @C=1em @R=.7em {
  \lstick{c}     &\qw   & \link{1}{-1}  & \qw                                              &\qw\link{1}{1}  &&   \ctrl 1         &\qw\\
  \lstick{q_1}   &\qw   & \link{-1}{-1}  & \multigate{3}{\text{\lstinline|rotations |}(m)}&\qw\link{-1}{1} &&   \gate{R_{m-n}} &\qw\\
  \lstick{q_2}   &\qw   & \qw             & \ghost{\text{\lstinline|rotations |}(m)}       &\qw          &\qw&\qw             &\qw\\
  \lstick{\vdots\ \ } & &            & \pureghost{\text{\lstinline|rotations |}(m)}       &             &&                   &\\
  \lstick{q_n}   &\qw   & \qw             & \ghost{\text{\lstinline|rotations |}(m)}       &\qw          &\qw&\qw             &\qw
 }}
\]

\bigskip
This standard QFT algorithm leaves the list in reverse order, and so for many purposes this program must be composed with a standard list reversal program, omitted here.

Here we have specified the QFT algorithm using recursion, in a way that is familiar from basic functional programming.
There are at least two drawbacks to this approach: (1) it is a priori unclear that the \lstinline|unbox fourier qs| terminates in general; (2) the type system does not tell us that (\lstinline|unbox fourier qs|) has the same length as \lstinline|qs|.
This is a familiar problem for list programming in a functional language such as Haskell, but it is particularly inconvenient for a quantum circuit layout engine, especially if one want to ensure that the terms that correspond to circuits are terminating and have a known size.
A good way to deal with it would be through some kind of dependent types~\cite[Sec.~6.2]{qwire}, for instance allowing a type $\mathrm{QList}(n)$ of arrays of qubits of size $n$ and
\[
\mathrm{fourier}:(n:\mathrm{Nat})\to \mathrm{Circ}(\mathrm{QList}(n),\mathrm{QList}(n))
\]

Integrating dependent types to EWire and associating such types to an appropriate categorical semantics is left for future work. Although this section discusses the integration of recursive types and how it leads to more refined presentations (in EWire) of quantum programs such as QFT, we do not provide a detailed explanation of the syntactic and semantic integration of recursive types. However, in a sequel of this paper co-authored by one of the authors of the present work~\cite{muqpl}, there is a detailed treatment of inductive types in quantum programming, both syntactically and semantically, with a computationally adequate denotational model based on W*-algebras.



\section{Relation to linear-non-linear models}%
\label{sec:LNL}

Recent work by Selinger and collaborators~\cite{rios-selinger-qpl17,ross-phd-thesis,malherbe-scott-selinger} has suggested that a good setting for quantum programming might be the linear/non-linear (LNL) models of linear logic in the style of Benton~\cite{benton}. This is, at first glance, a different approach to the QWire/EWire approach, for the following reason. In QWire/EWire the type of circuits $\Circ(W,W')$ is
a type in the host language, and is not itself a wire type. By contrast, with LNL models, the linear function space $(W\multimap W')$ is itself an object of the category
containing the linear types.

Rios and Selinger~\cite{rios-selinger-qpl17} gave a particular construction of an LNL model for quantum programming.
Mislove et al.~\cite{lindenhovius-mislove-zamdzhiev} have also been working on some related constructions.
In this section we propose a variation on that construction that works for a broad class of EWire models. In this way we bring the two lines of research closer together.

\begin{defiC}[\cite{benton}]
A \emph{linear/non-linear (LNL) model} is given by the following data:
\begin{enumerate}
 \item\label{item:lnl1} a symmetric monoidal closed category $\cat{L}$
 \item\label{item:lnl2} a cartesian closed category $\cat{H}$
 \item\label{item:lnl3} a symmetric monoidal adjunction $F \dashv G$ formed from symmetric monoidal functors $F:\cat{H}\to\cat{L}$ and $G:\cat{L}\to\cat{H}$.
\end{enumerate}
\end{defiC}

\noindent
There is an equivalent formulation of LNL models using enriched category theory. This plays a central role in work
on the enriched effect calculus~\cite{enriched-effect-calculus}.


\begin{lem}\label{lemma:enriched-lnl}
Let $\cat H$ be a cartesian closed category and $\cat L$ be a symmetric monoidal closed category.
The following data are equivalent:
\begin{enumerate}
\item\label{item:enriched-lnl1} structure making $\cat L$ an $\cat H$-enriched symmetric monoidal closed category with copowers;
\item\label{item:enriched-lnl2} a symmetric monoidal adjunction between $\cat H$ and $\cat L$ (i.e.~a LNL model).
\end{enumerate}
\end{lem}

\begin{proof}
From~(\ref{item:enriched-lnl1}) to~(\ref{item:enriched-lnl2}), let the adjunction be the copower/hom adjunction
\[
(-)\odot I\dashv \cat L(I,-):\cat (\cat{L},\otimes)\to \cat (\cat{H},\times)\text.
\]
which is always a symmetric monoidal adjunction.

Since $\cat L$ is symmetric monoidal closed,
each $-\otimes Z:\cat L\to\cat L$ has a right adjoint and so preserves weighted colimits, in particular copowers:
\[
((X\odot W)\otimes Z)\cong X\odot (W\otimes Z)
\]
In particular,
\begin{align*}
F(X\times Y)=(X\times Y)\odot I&=X\odot (Y\odot I)=X\odot (I\otimes (Y\odot I))
\\&=(X\odot I)\otimes (Y\odot I)=F(X)\otimes F(Y)\text.
\end{align*}
as required.

From~(\ref{item:enriched-lnl2}) to~(\ref{item:enriched-lnl1}), let the enrichment be given by
\[
\cat L(X,Y)\defeq G(X\multimap Y)\text.
\]
Then $\cat L$ has copowers given by $h\odot X\defeq F(h)\otimes X$.
For
\begin{align*}
\cat H(h'\times h,G(C\multimap D))
&\cong
\cat C(F(h'\times h),C\multimap D)
\cong
\cat C(F(h')\otimes F(h),C\multimap D)
\\&\cong
\cat C(F(h'),(F(h)\otimes C)\multimap D))
\cong
\cat H(h',G(F(h)\otimes C)\multimap D)
\end{align*}

One must show that the symmetric monoidal closed structure is enriched. For example, we must give a map
\[
G(C\multimap C')\times G(D\multimap D')\to G(C\otimes C' \multimap D\otimes D')
\]
To give such a map is to give a map
\[
FG(C\multimap C') \otimes FG(D\multimap D')\to (C\otimes C')\multimap (D\otimes D')
\]
and we use the one that arises from the counit of the adjunction.

Finally we show that passing from~(\ref{item:enriched-lnl1}) to~(\ref{item:enriched-lnl2}) and back to~(\ref{item:enriched-lnl1}) gives the same enriched structure up to isomorphism. But this is trivial, because
\[
\cat L(I,X\multimap Y)\cong \cat L(X,Y)
\]
And we must show that passing from~(\ref{item:enriched-lnl2}) to~(\ref{item:enriched-lnl1}) and back to~(\ref{item:enriched-lnl2}) gives the same adjunction, but this is also fairly trivial because
$F$, being a monoidal left adjoint, must preserve copowers and the monoidal structure:
\[F(X)\cong F(X\times 1) \cong X\odot F(1)\cong X\odot I\text. \qedhere\]
\end{proof}

\newcommand{\LNLwire}{LNL-EWire\xspace}
\begin{defi}
An \emph{\LNLwire model} $(\cat L, \cat H, \cat{L}_0, \cat{H}_0, F, G)$ is given by a linear/non-linear model $(\cat L,\cat H,F,G)$ together with
a small symmetric monoidal full subcategory $\cat L_0\subseteq \cat L$ and a small full subcategory
$\cat H_0\subseteq \cat H$ such that $F$ restricts to a functor $\cat H_0\to \cat L_0$.
\end{defi}

\begin{prop}%
\label{prop:Lwire-to-Ewire}
Every \LNLwire model $(\cat L, \cat H, \cat{L}_0, \cat{H}_0, F, G)$ induces an EWire model, when we put
\begin{itemize}
\item $\cat C=\cat L_0$;
\item The enrichment of $\cat C$ is given by $\cat C(X,Y)=G(\cat L(X,Y))$;
\item $T=GF$.
\end{itemize}
\end{prop}
\begin{proof}
Let us verify that the axioms of Definition~\ref{def:model-QWire} are verified for the quadruplet $(\cat{L}_0,\cat{H}_0,\cat H,GF)$.

Axiom~(\ref{enum:interp-circuits}) follows from Lemma~\ref{lemma:enriched-lnl}.
For Axiom~(\ref{enum:copow}) we need to show that $\cat L_0$ has copowers by objects of $\cat H_0$.
Lemma~\ref{lemma:enriched-lnl} already gives us that $\cat L$ has copowers,
and so we must show that a copower $h\odot X$ is in $\cat L_0$ when $h\in\cat H_0$ and $X\in\cat L_0$.
First, we observe that in $\cat L$, the following equation holds, because $(-)\otimes X$ preserves copowers since it a right adjoint.
\begin{equation}
\label{eq:copowers}
(h\odot X)\cong (h\odot (I\otimes X))\cong (h\odot I)\otimes X
\end{equation}
\LNLwire models are defined to be such that $F$ restricts to a functor $\cat H_0\to \cat L_0$.
We know that $F\cong ((-)\odot I)$, so $(h\odot I)$ is in $\cat L_0$.
It follows that $h\odot X$ is in $\cat L_0$ because, in an \LNLwire model, $\cat L_0$ is a symmetric monoidal subcategory.

For Axiom~(\ref{enum:SMC-copow}), we must show that $A\otimes -$ preserves copowers.
This again follows from $A\otimes -$ having a right adjoint, the closed structure, and hence preserving colimits.

For Axiom~(\ref{enum:monad}) we must give an enriched relative monad morphism,
but by Lemma~\ref{lemma:enriched-lnl} this can be the identity morphism.
\end{proof}

\begin{thm}%
\label{th:Ewire-to-Lwire}
For every EWire model $(\cat C,\cat H,\cat H_0,T)$ there is an \LNLwire model
\[(\cat L,\cat H,\cat L_0,\cat H_0,F,G)\] with a $\cat L_0=\cat C$ and
a relative monad morphism $GFj\to T$.
\end{thm}

The proof of this theorem is postponed until after the following lemma, which is hinted in~\cite{kelly} and is
likely well known in some circles.

\newcommand{\colclass}{\mathcal{F}}
\begin{lem}\label{lemma:colimitcompletion}
Let $\cat C$ be a symmetric monoidal category enriched in a cartesian closed category $\cat H$, and suppose that $\cat H$ is also locally presentable as an enriched category.
Let $\mathcal F$ be a class of weighted colimits in $\cat C$.
Let $\bar {\cat C}\defeq {[\cat C\op ,\cat H]}_{\mathcal F}$ be the $\cat H$-category of
$\colclass$-limit preserving $\cat H$-functors.
\begin{enumerate}
\item\label{item:colim1} The restriction of the Yoneda embedding $\cat C\to\bar{\cat C}$
preserves $\colclass$-colimits, and
exhibits the free $\cat H$-colimit cocompletion of $\cat C$
as a category with $\colclass$-colimits.
\item\label{item:colim2} If $\colclass$-colimits distribute over the monoidal structure, in that the canonical maps
\[
\colim^W_{j\in J}(C_j\otimes D)\ \longrightarrow\
(\colim^W_{j\in J}C_j)\otimes D
\]
are isomorphisms, then $\bar {\cat C}$
admits a symmetric monoidal closed structure,
and the Yoneda embedding $\cat C\to \bar {\cat C}$ strongly preserves the symmetric monoidal structure.
\end{enumerate}
\end{lem}

\begin{proof}
For (\ref{item:colim1}), we first recall that it is straightforward to show that the Yoneda embedding factors through $\cat C\to\bar{\cat C}$, this follows from the definition of limit. Moreover, it is routine to check that this restricted Yoneda embedding preserves $\colclass$-colimits: this follows from the Yoneda lemma.

Less trivial is the fact that the category $\bar{\cat C}$ of $\colclass$-limit preserving functors is a reflective subcategory of the category of all functors $\hat{\cat C}$, i.e.\ that the embedding $\bar{\cat C}\to\hat {\cat C}$ has a left adjoint. This can be proved using a general method of orthogonality subcategories (e.g.~\cite[Ch.~6]{kelly}), since a functor is in $\bar {\cat C}$ if it is
orthogonal to the canonical morphism
\[
\colim^W_i \cat C(-,d_i)\to \cat C(-,\colim^W_i d_i)
\]
for each diagram in the class $\colclass$. From this reflection property we conclude that $\bar{\cat C}$ has all colimits.

Now we consider an $\colclass$-colimit preserving functor $F:\cat C\to \cat D$ where $\cat D$ is cocomplete.
We show that $F$ extends essentially uniquely to a functor $F_!:\bar{\cat C}\to \cat D$ that preserves all colimits.
\[
\xymatrix{\cat C\ar[drr]_F\ar[rr]^{\cat C(=,-)}&&\bar{\cat C}\ar@{..>}[d]^{F_!}\\&&\cat{D}} 
\]
First recall that $F$ already extends essentially uniquely to a functor $F_!:\hat{\cat C}\to \cat D$,
which is right adjoint to the functor $\cat D(F(=),-): \cat D\to \hat {\cat C}$.
Moreover $\cat D(F(=),-)$ factors through $\bar {\cat C}\subseteq \hat {\cat C}$:  this follows from the assumption that $F$ preserves
$\colclass$-colimits.

So the adjunction
\[
F_!\dashv \cat D(F(=),-): \cat D\to \hat {\cat C}
\]
restricts to an adjunction
\[
F_!\dashv \cat D(F(=),-): \cat D\to \bar {\cat C}
\]
and so $F_!:\bar{\cat C}\to \cat D$ preserves colimits, and is the required extension of $F$.

This functor $F_!:\bar{\cat C}\to \cat D$ is essentially unique as a colimit preserving functor such that $F_!\circ \cat C(=,-) \cong F$.
Indeed suppose that $L:\bar{\cat C}\to \cat D$ preserves colimits and $L\circ \cat C(=,-)\cong F$.
Since $\bar{\cat C}$ is a reflective subcategory of a presheaf category, it is `total' hence $L$ has a right adjoint $R:\cat D\to \bar{\cat C}$.
By the Yoneda lemma,
\[R(d)(c)\cong \bar{\cat C}(\cat C(-,c),R(d))\cong \cat D(L(\cat C(-,c)),d)\cong \cat D(F( c),d)\]
and so, by uniqueness of adjoints, $L\cong F_{!}$.

As for~(\ref{item:colim2}), recall that the presheaf category $\hat{\cat C}$ has a symmetric monoidal closed such that the
Yoneda embedding $\cat C\to \hat{\cat C}$ is strongly monoidal.
This structure is due to Day~\cite{day-convolution}, and, for $P,Q\in\hat {\cat C}$, we have
\begin{align*}&(P \Day Q) (Z)= \textstyle{\int^{X,Y}
    \cat{C}(X \otimes Y, Z) \times P(X) \times Q(Y)}\\
&(P\multimap Q)(X)=\hat{\cat C}(P,Q(X\otimes -))
\end{align*}
If $Q$ preserves $\colclass$-limits, then, because each $(X\otimes-):\cat C\to \cat C$ preserves $\colclass$-limits too,
$(P\multimap Q)$ preserves $\colclass$-limits. In other words, if $Q\in\bar{\cat C}$ then $(P\multimap Q)\in\bar {\cat C}$.
This condition is sufficient for extending $(\multimap)$ to a symmetric monoidal closed structure on $\bar {\cat C}\subseteq \hat{\cat C}$, such that the reflection $\hat {\cat C}\to \bar{\cat C}$ is strongly monoidal~\cite{day-reflection}.
(Note that this does not mean that the embedding $\bar{\cat C}\to \hat {\cat C}$ is strongly monoidal.)
\end{proof}

\begin{proof} (Proof of Theorem~\ref{th:Ewire-to-Lwire}.)
Let $\colclass$ be the class of copowers by objects of $\cat H_0$.
Let $\cat L \defeq {[{\cat C}\op,\cat H]}_{\colclass}$
be the category of $\cat H_0$-power preserving
$\cat H$-functors.
By Lemma~\ref{lemma:colimitcompletion}, the category $\cat L$ has a symmetric monoidal closed structure, and the Yoneda embedding $\cat C\to \cat L$ preserves it.
Let $\cat L_0\defeq \cat C$.
By Lemma~\ref{lemma:enriched-lnl}, this structure gives rise to an LNL model.

Finally, we must give a relative monad morphism $GFj\to T$.
Recall that we assume that in an \LNLwire model, $F$ restricts to a functor $\cat H_0\to \cat L_0=\cat C$.
By Lemma~\ref{lemma:enriched-lnl}, $F=((-)\odot I)$,
and this restriction property means that $\cat L_0$ is closed under copowers by objects of $\cat H_0$.
So $GFj=\cat L(I,j(-)\odot I)\cong \cat C(I,j(-)\odot I)$, and the relative monad morphism
is the $(\textrm{run})$ structure of the EWire model.
\end{proof}

From Theorem~\ref{th:Ewire-to-Lwire}, we deduce that every EWire model $(\cat{C},\cat{H},\cat{H}_0,T)$ induces an exponential construction $! \defeq J \circ \cat{L}(I,-): \cat L \to \cat L$ where
$\cat L \defeq {[{\cat C}\op,\cat H]}_{\colclass}$.
\hide{The increasing prevalence of the use of presheaves in the study of the semantics of quantum programming languages led us on a path to the development of a quantum domain theory, sketched in Section~\ref{sec:B3:quantum-domain}.}
For example, starting from the EWire model
of Proposition~\ref{prop:B2:fdcpu-model}
\[(\FdCPU^{\mathrm{op}},\cat{Set},\N,\DM)\]
we arrive at an LNL-EWire model
\[\Big({[\FdCPU,\cat{Set}]}_{\mathcal F},\cat{Set},\FdCPU^{\mathrm{op}},\N,F,G\Big)\]
where $\mathcal F$ is the class of finite products,
and where
$G:{[\FdCPU,\cat{Set}]}_{\mathcal F}$ is given by
$G(P)=P(\mathbb C)$.
This is close to the LNL model for Proto-Quipper proposed in~\cite{rios-selinger-qpl17}.

Starting from the EWire model of (\ref{eqn:cpudcpo}),
\[(\FdCPU^{\mathrm{op}},\cat{Dcpo},\N,\mathcal V)\]
we arrive at an LNL-EWire model
\[\Big({[\FdCPU,\cat{Dcpo}]}_{\mathcal F},\cat{Dcpo},\FdCPU^{\mathrm{op}},\N,F,G\Big)\]
where $\mathcal F$ is again the class of finite products.
Starting from the EWire model of subunital maps (\ref{eqn:cpsudcpo}),
\[(\Fd\CPSU^{\mathrm{op}},\cat{Dcpo},\N,\mathcal V)\]
we arrive at an LNL-EWire model
\[\Big({[\Fd\CPSU,\cat{Dcpo}]}_{\mathcal F},\cat{Dcpo},\Fd\CPSU^{\mathrm{op}},\N,F,G\Big)\]
where now ${[\Fd\CPSU,\cat{Dcpo}]}_{\mathcal F}$ comprises locally continuous product preserving functors. These models are
related to the LNL model for recursion proposed by~\cite{lindenhovius-mislove-zamdzhiev}.
Indeed, presheaf models have been proposed for quantum programming by various authors, perhaps beginning from~\cite{malherbe-scott-selinger}.

\subsection*{Summary}

Having described a family of embedded languages for (quantum) circuits in Section~\ref{sec:qwire}, we described their denotational models in enriched category theory in Section~\ref{sec:cat-qwire}.
In Section~\ref{sec:ewire-domains}, after an exploration of EWire models enriched over dcpos, we explored some of the possible extensions of the languages considered in this work.
Finally in Section~\ref{sec:LNL}, we established a connection with the Benton's notion of linear-non-linear models.

\subsection*{Acknowledgements}

The authors would like to thank Tom Hirschowitz, Bart Jacobs, Shane Mansfield, Paul-Andr\'e Melli\`es, Michele Pagani, and Peter Selinger for helpful discussions, but also Bert Lindenhovius, Michael Mislove, and Vladimir Zamdzhiev for hosting the first author at Tulane University during the early stage of the elaboration of this work, and Jennifer Paykin and Robert Rand for introducing us to the subtle aspects of QWire.
The research leading to these results has received funding from the ERC grant agreement n.~320571, a Royal Society fellowship, and the EPSRC grant EP/N007387/1.

\bibliographystyle{plain}
\bibliography{mfps33-lmcs}

\end{document}